\documentclass[twocolumn]{aastex701}
\shorttitle{SHIELD UV SFRs}
\shortauthors{Gormanous et al.}

\begin{document}

\title{UV Star-Formation Rates of the SHIELD Dwarf Galaxies }

\correspondingauthor{John Salzer}
\email{josalzer@iu.edu}

\author[0009-0001-1540-3246]{David G. Gormanous}
\affiliation{Department of Astronomy, Indiana University, 727 East Third Street, Bloomington, IN 47405, USA}
\email{dgormano@iu.edu}

\author{Anjali S. Dziarski}
\affiliation{Department of Astronomy, Indiana University, 727 East Third Street, Bloomington, IN 47405, USA}
\email{anjdziar@alumni.iu.edu}

\author[0000-0001-8483-603X]{John J. Salzer}
\affiliation{Department of Astronomy, Indiana University, 727 East Third Street, Bloomington, IN 47405, USA}
\email{josalzer@iu.edu}

\author[0000-0002-1821-7019]{John M. Cannon}
\affiliation{Department of Physics and Astronomy, Macalester College, Saint Paul, MN 55105, USA}
\email{jcannon@macalester.edu}

\author[0000-0001-9165-8905]{Steven Janowiecki}
\affiliation{Hobby-Eberly Telescope, McDonald Observatory, University of Texas, Austin, TX 78712, USA}
\email{steven.janowiecki@gmail.com}

\author[0000-0001-5247-1371]{Nathalie Haurberg}
\affiliation{Department of Physics and Astronomy, Knox College, 2 East South Street, Galesburg, IL 61401, USA}
\email{nhaurber@knox.edu}

\author[0009-0005-3076-1104]{April Horton}
\affiliation{Department of Physics and Astronomy, Macalester College, Saint Paul, MN 55105, USA}
\affiliation{Department of Physics and Astronomy, Texas Christian University, Fort Worth, TX 76129, USA}
\email{aprhorton@gmail.com}

\author[0000-0002-9798-5111]{Elizabeth A. K. Adams}
\affiliation{ASTRON, The Netherlands Institute for Radio Astronomy, Oude Hoogeveensedijk 4, 7991 PD, Dwingeloo, The Netherlands}
\affiliation{Kapteyn Astronomical Institute, University of Groningen, Postbus 800, 9700 AV Groningen, The Netherlands} 
\email{adams@astron.nl}

\author[0000-0001-5334-5166]{Martha P. Haynes}
\affiliation{Center for Astrophysics and Planetary Science, Space Sciences Building, Cornell University, Ithaca, NY 14853, USA}
\email{haynes@astro.cornell.edu}

\author{Myles J. Klapkowski}
\affiliation{Department of Physics and Astronomy, Macalester College, Saint Paul, MN 55105, USA}
\email{mklapkowski12@gmail.com}

\author{Joshua R. Marine}
\affiliation{Department of Physics and Astronomy, Macalester College, Saint Paul, MN 55105, USA}
\email{jmarine@macalester.edu}

\author[0000-0001-5538-2614]{Kristen B. W. McQuinn}
\affiliation{Space Telescope Science Institute, 3700 San Martin Drive, Baltimore, MD 21218, USA}
\email{kmcquinn@stsci.edu}

\author[0000-0001-8283-4591]{Katherine L. Rhode}
\affiliation{Department of Astronomy, Indiana University, 727 East Third Street, Bloomington, IN 47405, USA}
\email{krhode@iu.edu}

\author[0000-0003-0605-8732]{Evan D. Skillman}
\affiliation{Minnesota Institute for Astrophysics, School of Physics and Astronomy, University of Minnesota, 116 Church Street, S.E., Minneapolis, MN 55455, USA}
\email{skill001@umn.edu}

%\author[]{}
%\affiliation{}

%\author{more TBD...}

\begin{abstract}

The Survey of \ion{H}{1} in Extremely Low-mass Dwarfs (SHIELD) is a  multi-wavelength observational project targeting gas-rich, star-forming dwarf galaxies at the faint end of the \ion{H}{1} mass function.  We present near-ultraviolet (NUV) and far-ultraviolet (FUV) flux measurements obtained from GALEX survey images and use these fluxes to derive FUV star-formation rates (SFRs) for all 75 SHIELD galaxies with GALEX data. This paper represents the first published analysis that makes use of the full SHIELD sample. We compare the FUV SFRs to a variety of physical quantities to better understand the nature of SHIELD galaxies. When comparing the H$\alpha$ SFRs to the FUV SFRs for SHIELD and other local galaxy surveys, we confirm and solidify previous results that show that H$\alpha$-based SFRs for dwarf galaxies can grossly underestimate the true rate of star formation, emphasizing the highly stochastic nature of star formation in extremely low-mass galaxies.  We further show that FUV SFRs appear to be robust tracers of star formation down to the very lowest galaxy masses included in this study.  We show that baryonic mass is the best mass-related tracer for the prediction of the FUV SFR of the galaxies in our sample.  Not surprisingly, the SHIELD dwarf galaxies exhibit long gas-depletion timescales and large gas mass fractions: fully 68\% of the SHIELD galaxies have gas masses that are larger than their stellar masses.

\end{abstract}

%\keywords{TBD}


%************************************************************
\section{Introduction} \label{sec:intro}

The study of dwarf galaxies is important for improving our understanding of the various physical processes that shape the evolution of galaxies.  Dwarf galaxies are more common in the universe than massive galaxies, and when surveyed to adequate depths they can yield robust samples in the nearby universe to study various galactic characteristics and evolutionary processes.  At the same time, the smaller size scales and relatively simple internal kinematics of dwarf galaxies allow them to be accessible laboratories for studying key processes such as star formation, feedback, and chemical evolution, which are crucial for our overall understanding of galaxy evolution. 

The process by which galaxies convert their gas into stars is perhaps the most important driver of their evolution.  Numerous studies have been carried out to investigate the star-formation rates (SFRs) of dwarf galaxies \citep[e.g.,][]{Lee_2009, Walter_2008, hunter2012, Roychowdhury_2014, teich2016, ott2012}.  One result from these studies has been that the star-formation efficiencies and the ability to form high-mass stars appears to be different for systems with lower SFRs typical of dwarf galaxies compared to higher-mass spirals.  Despite the insights generated by these and other studies, a comprehensive understanding of the star-formation process in dwarf galaxies has yet to be established.  Additional focus on the processes that turn gas into stars in low-mass galaxies provides interesting avenues for improving our knowledge of the evolution of galaxies in general.

The goal of our current project is to advance our knowledge of how star formation proceeds in dwarf galaxies by studying the properties of a unique sample of dwarf galaxies cataloged in the Survey of \ion{H}{1} in Extremely Low-mass Dwarfs (SHIELD) \citep{cannon2011}.  The SHIELD galaxy sample differs from those used in previous studies in two important ways.  First, it represents a {\it complete sample} of the lowest \ion{H}{1} mass galaxies within the local universe, allowing us to probe the star-formation characteristics at the extreme low-mass end of the galaxy {\ion{H}{1}} mass function.  Second, the selection method used to construct the SHIELD sample means that our sample is {\it comprehensive} in the sense that it is not biased by the optical characteristics of the galaxies (see below). 

The focus of the present paper is to explore the FUV SFRs of the full SHIELD sample.  This work expands upon an earlier study that looked at the FUV SFRs of the original sample of 12 SHIELD galaxies \citep{teich2016} by carrying out a similar analysis and presenting results for the full sample of 82 SHIELD galaxies.   A companion paper will present our narrowband images and derived H$\alpha$ SFRs for the complete SHIELD sample (Shepley et al. 2026, in preparation).

This paper is laid out as follows: Section~\ref{sec:shield} will introduce the SHIELD galaxy sample and give a summary of previous SHIELD-based studies. Section~\ref{sec:sfrdet} will describe the methodology for our UV flux measurements and SFR calculations.  Section~\ref{sec:sfrrel} will present the results of our FUV SFR comparisons, highlighting the properties of both the SHIELD galaxies and an extensive comparison sample. Section~\ref{sec:summary} will summarize the main results of this paper.
%Section~\ref{sec:Discussion} will discuss our results compared to previous findings for our galaxy sample, 

% I'm referencing Section~\ref{sec:intro}

%************************************************************
\section{The SHIELD Sample of Dwarf Galaxies} \label{sec:shield}

SHIELD is an ongoing project that aims to refine our understanding of the properties of a comprehensive sample of \ion{H}{1}-selected dwarf galaxies in the local universe. The initial SHIELD sample \citep{cannon2011} consisted of 12 sources from the early portions of the Arecibo Legacy Fast ALFA (ALFALFA) survey \citep{giovanelli2005}.  The sample was  enlarged to 30 galaxies when ALFALFA had surveyed 40\% of the sky \citep{haynes2011} and was subsequently expanded to 82 galaxies once the ALFALFA survey was complete \citep{haynes2018}. The full SHIELD catalog is a volumetrically complete sample of the sky within the given selection constraints listed below.  The ALFALFA-detected galaxies populate the low-mass end (e.g., M$_{HI}$ $<$ 10$^7$ M$_\odot$) of the \ion{H}{1} mass function at a significantly improved level compared to previous \ion{H}{1} surveys \citep{martin2010, jones2018}, making this survey essential to our understanding of the properties of the lowest-mass dwarf galaxies.

The primary selection criteria of the SHIELD sample include: (i) \ion{H}{1} gas mass below 1.6 $\times$ 10$^7$ M$_\odot$, (ii) \ion{H}{1} velocity widths (FWHM) of below 65 km s$^{-1}$ (to exclude massive but gas-poor galaxies),  (iii) the presence of a detected stellar population as indicated by stellar emission visible in existing wide-field imaging surveys such as SDSS, and (iv) distances less than $\sim$11 Mpc to ensure that the galaxies were close enough for detailed analysis \citep{cannon2011, mcquinn2021}.  In general, nearby galaxies (distances less than $\sim$4 Mpc) that satisfied these selection criteria but were already included in existing \ion{H}{1} surveys (e.g., LITTLE THINGS, FIGGS, VLA ANGST; see references below) were not included in SHIELD.  Hence, SHIELD is {\it complete} in the sense that it includes all dwarf galaxies detected by ALFALFA with M$_{HI}$ below 1.6 $\times$ 10$^7$ M$_\odot$ within the ALFALFA survey area and with distances in the range of approximately 4 to 11 Mpc but excluding a handful of previously known and well-studied galaxies.

We refer above to the fact that the SHIELD sample is {\it comprehensive}.  By this we mean that the selection criteria do not bias the properties of the galaxies unduly, with the exception that they have a measurable cold gas content.  This results in a sample of gas-bearing galaxies that is neither biased toward high optical surface brightness (HSB) and high SFRs nor toward low optical surface brightness (LSB).  SHIELD includes a broad mixture of both HSB and LSB galaxies.  The requirement that the SHIELD galaxies contain a detectable amount of \ion{H}{1} gas does mean that they tend to show evidence for at least some recent star formation (e.g., within the past 100-200 Myr); their optical colors tend to be rather blue \citep[e.g.,][]{cannon2011}.   However, only a few would be categorized as exhibiting strong current star formation (e.g., a blue compact dwarf; see Section~\ref{sec:StellarsSFR}) %\S 4.5).

A key aspect of the SHIELD project is the study of dwarf galaxies over a range of wavelengths to better understand their essential characteristics, including their gaseous content, stellar constituents, star formation properties, and the state of their chemical evolution.  Current efforts are focused on obtaining complete and uniform data sets for all 82 galaxies.  Examples of these ancillary data sets include ground-based optical broadband and narrowband (H$\alpha$) imaging, optical spectroscopy of embedded \ion{H}{2} regions, multi-configuration VLA and WSRT \ion{H}{1} synthesis imaging, Spitzer 3.6 $\mu$m NIR imaging, and GALEX NUV and FUV fluxes.  In addition, approximately half of the SHIELD galaxies have been imaged with HST and have secure distances from TRGB measurements.

Early studies of the original 12 galaxies have already provided a preliminary glimpse of the characteristics of the SHIELD dwarfs.  These studies include \citet{cannon2011} (\ion{H}{1} distributions and kinematics), \citet{mcquinn2014} (HST images and TRGB distances), \citet{haurberg2015} (H$\alpha$ imaging, spectroscopy, and metallicities), \citet{mcquinn2015a} (star-formation histories), \citet{teich2016} (SFRs and star formation efficiencies), and \citet{mcnichols2016} (\ion{H}{1} kinematics and dynamical masses).  More recently, \citet{mcquinn2021} has added additional HST images, TRGB distances, star-formation histories, and gas kinematics for the intermediate-sized sample of 30 SHIELD galaxies.  Finally, several SHIELD galaxies were used to explore the shape of the low-mass end of the Baryonic Tully-Fisher relation \citep{mcquinn2022}.

%************************************************************

\section{UV Flux Measurements and Star-Formation Rates}\label{sec:sfrdet}

The SHIELD project uses both H$\alpha$ narrowband and FUV emission data to estimate SFRs. The H$\alpha$ emission traces \ion{H}{2} regions ionized by young, massive OB stars, enabling estimates of very recent star formation. H$\alpha$-based SFRs are often referred to as ``current" or ``instantaneous" SFRs \citep{Kennicutt1998a}, because they are sensitive to star formation on a characteristic timescale of $\sim$10$^{}$ Myr.  H$\alpha$ images for the full SHIELD sample have been obtained, and SFRs have been measured for all 55 galaxies with detectable H$\alpha$ emission \citep[][Shepley et al. 2026, in preparation]{shepley_2020}.  UV images detect the radiation from young, bright stars, which have characteristic spectral types of B and A, and characteristic lifetimes of 100-200 Myr. Hence, UV-based SFRs are measuring the star-forming activity in galaxies over a longer timescale than those derived from H$\alpha$ observations.

\subsection{Measuring UV Fluxes from GALEX}\label{sec:flux}
GALEX \citep{martin2005} survey images were used to obtain UV flux measurements for the SHIELD galaxies. Measurements of NUV and FUV fluxes were carried out on background-subtracted GALEX images using an interactive photometry script which allowed the user to mask any intervening objects and measure the total UV flux.  %utilized a user-supplied aperture size to determine the galaxy's flux and mask any intervening objects.  
GALEX data are available for 76 of the 82 SHIELD galaxies.  The remaining 6 galaxies were not included in any GALEX images (AGC 748779, AGC 112503, AGC 200232, AGC 238890, AGC 215282, and AGC 229052).  One galaxy (AGC 124056) was too faint in the GALEX images to allow for a reliable measurement of its NUV and FUV fluxes.  

\begin{figure}
\centering
\includegraphics[width=3.35in]
{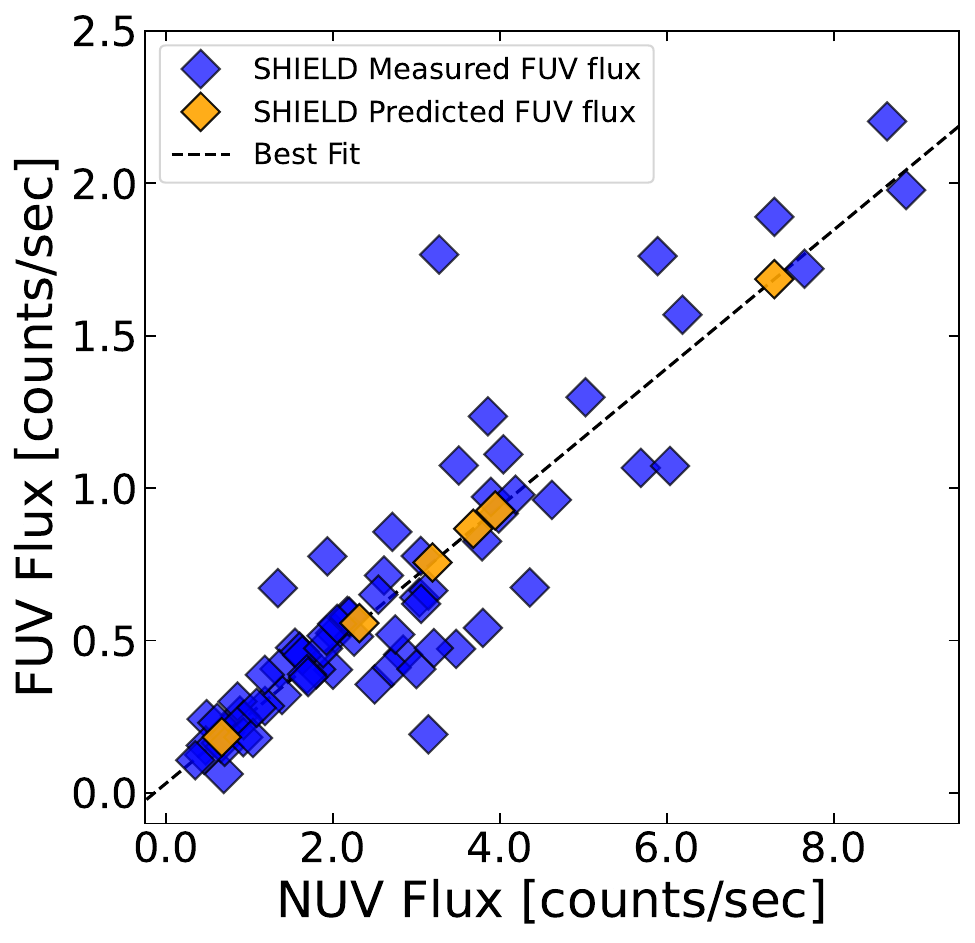}
\caption{Graph of FUV flux vs.\ NUV flux for the SHIELD galaxies in units of counts s$^{-1}$. The blue diamonds indicate galaxies with measured FUV fluxes and the orange diamonds indicate galaxies with predicted FUV fluxes. The dashed line indicates a best fit line with a slope of 0.227. The scatter about the best fit line is 0.21 counts s$^{-1}$.}
\label{fig:fluxcomp}
\end{figure}

We produced NUV flux measurements for 75 galaxies and FUV flux measurements for 69 galaxies. The 6 galaxies missing FUV fluxes were not observed by GALEX through the FUV channel, presumably because these fields were not observed prior to the 2009 failure of the FUV detector.
%and one galaxy (AGC 124056) was too faint in the GALEX images to accurately measure its NUV and FUV flux. 
We constructed a relationship comparing FUV and NUV fluxes for galaxies with both.  This relationship was well fit by a straight line which had a scatter of 0.21 counts s$^{-1}$. This was used to predict the FUV fluxes of the 6 galaxies missing FUV data with the following equation:
\begin{equation}
  FUV=0.227 \times NUV+0.032,
\end{equation}
where the FUV flux and the NUV flux are in counts s$^{-1}$. The plot of FUV flux vs.\ NUV flux is given in Figure~\ref{fig:fluxcomp}, which includes the points with predicted FUV values, as well as the best fit line from which we derived the equation to predict these FUV fluxes. The SHIELD galaxies with predicted FUV fluxes are noted in Table~\ref{tab:uvflux}.

\begin{figure*}
    \centering
    \vskip -0.07in
    \includegraphics[width=5.90in]{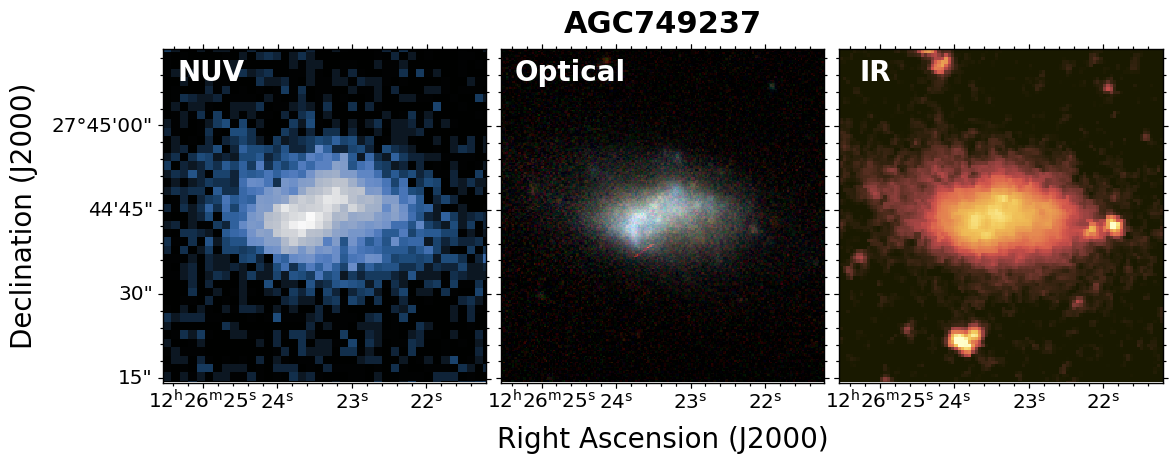}
    \vskip -0.27in
    \includegraphics[width=5.90in]{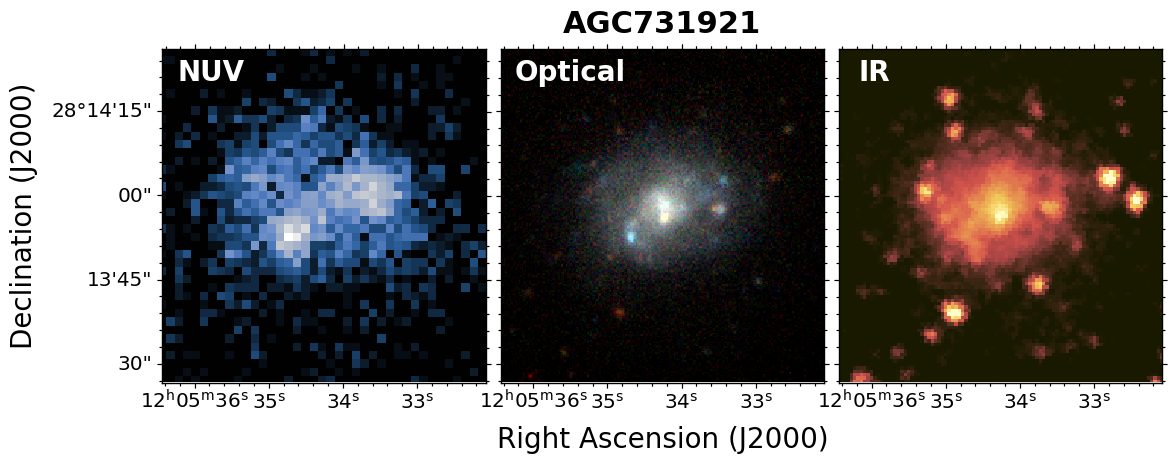}
    \vskip -0.27in
%    \includegraphics[width=4.80in]{AGC103722_NUV_opt_IR.png}
%    \vskip -0.22in
    \includegraphics[width=5.90in]{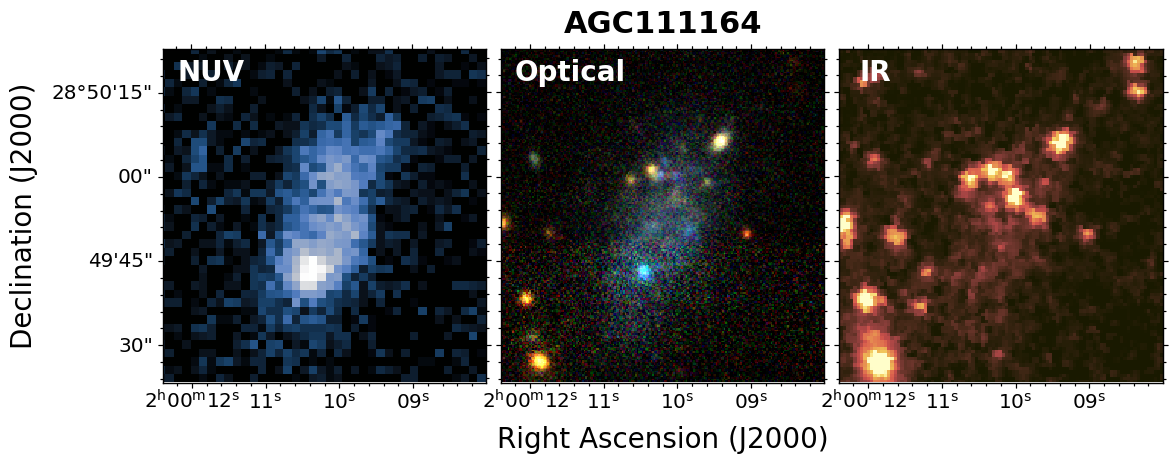}
    \vskip -0.27in
    \includegraphics[width=5.90in]{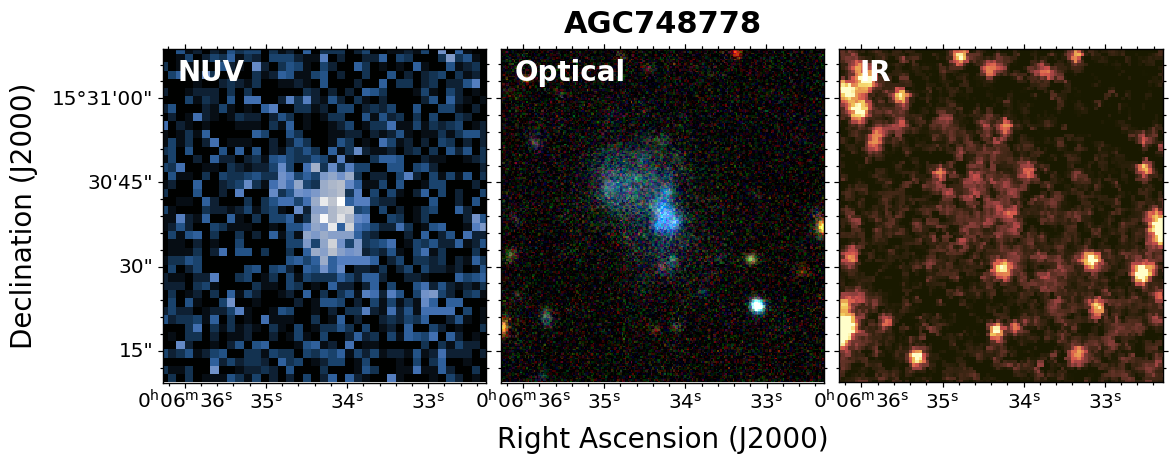}
    \caption{Example images for four representative SHIELD galaxies.  Each row contains three images of the same galaxy obtained in distinct wavelength regimes: NUV, optical, and NIR.  The individual image cutouts are each 1 by 1 arcmin on a side.  The four galaxies shown cover the full range of observed FUV SFR for the SHIELD sample, and are ordered by decreasing SFR: {\it top} -- AGC 749237, log(SFR$_{FUV}$) = $-$2.16; {\it 2nd row} -- AGC 731921, log(SFR$_{FUV}$) = $-$2.42; {\it 3rd row} -- AGC 111164, log(SFR$_{FUV}$) = $-$3.09; {\it bottom} -- AGC 748778, log(SFR$_{FUV}$) = $-$3.35.}
    \label{fig:examples}
\end{figure*}

In Figure~\ref{fig:examples} we present images of four representative SHIELD galaxies.  Each row of the figure consists of images of the same galaxy acquired at three different wavelengths.  The leftmost images show GALEX NUV data, the middle set of images are color-composite optical images (g, r, and z bands) from the DESI Legacy Imaging Surveys \citep{dey_2019}, and the rightmost set are 3.6 $\mu$m images obtained with the Spitzer Space telescope \citep{marine_2023}.  The GALEX NUV images are part of the dataset being analyzed in the current study, and were chosen for use in Figure~\ref{fig:examples} because they are deeper than the corresponding FUV images and better illustrate the UV morphologies.  Additional examples of GALEX UV images of SHIELD galaxies are shown in \citet{teich2016}.

%\startlongtable
%\centerwidetable
\begin{deluxetable*}{ccccccccccc}
\digitalasset
\tabletypesize{\scriptsize}
%\tablewidth{0pt} 
%\tablenum{1}
\tablecaption{Measured UV Fluxes for SHIELD Galaxies\label{tab:uvflux}}
\tablehead{
& & & &\underbar{\ \ \ \ \ \ \ \ \ \ \ \ \ \ \ }&\underbar{\ \ NUV\ \ }&\underbar{\ \ \ \ \ \ \ \ \ \ \ \ } &\underbar{\ \ \ \ \ \ \ \ \ \ \ \ \ \ \ }&\underbar{\ \ FUV\ \ }&\underbar{\ \ \ \ \ \ \ \ \ \ \ \ } & \\
\colhead{AGC \#} & \colhead{RA} & \colhead{Dec} & \colhead{E(B-V)} & \colhead{Exp. Time} & \colhead{Aper.} & \colhead{Flux} & \colhead{Exp. Time} & \colhead{Aper.} & \colhead{Flux} & Notes \\
& (J2000) & (J2000) & mag & sec & arcsec & cnts/sec & sec & arcsec & cnts/sec \\
(1) & (2) & (3) & (4) & (5) & (6) & (7) & (8) & (9) & (10) & (11) 
} 
\startdata 
102728 &  \ \ 0.0891667 & 31.0219440 &  0.046 &    1648 &    15 &   0.7074 &    1648 &    10 &   0.1984 \\
103722 &  \ \ 3.6916666 & 10.8130560 &  0.095 &      80 &    15 &   3.1916 & ... & --- &   0.7564 & 3 \\ %FUV flux computed \\
104208 & \ 10.4270830 & 12.9922228 &  0.100 &     112 &     8 &   0.4741 &     112 &     8 &   0.1560 \\
110482 & \ 25.5720844 & 26.3666668 &  0.091 &    1601 &    50 &   3.5069 &    1601 &    40 &   1.0744 \\
111164 & \ 30.0420837 & 28.8311119 &  0.055 &    2269 &    50 &   1.9311 &    2269 &    50 &   0.7764 \\
111946 & \ 26.6758327 & 26.8013878 &  0.082 &    3109 &    40 &   1.3385 &    1532 &    40 &   0.6724 \\
111977 & \ 28.8341675 & 27.9538898 &  0.070 &    1702 &    42 &   3.8559 &    1702 &    40 &   1.2357 \\
112503 & \ 24.5012493 & 14.9827776 &  0.055 & ... & ... &  ... & ... & ... &  ... & 1 \\%No GALEX data \\
112505 & \ 25.0400009 & 15.9400005 &  0.065 &    3197 &    13 &   0.6124 &    1693 &    13 &   0.2308 \\
112521 & \ 25.2833347 & 27.3222237 &  0.061 &    1233 &    40 &   0.4839 &    1233 &    30 &   0.2426 \\
\\
123352 & \ 42.1633339 & 23.2744446 &  0.234 &     112 &    15 &   1.9993 &     112 &    15 &   0.4048 \\
124056 & \ 44.4099998 & 23.8033333 &  0.162 & 109 & ... &  ... & 109 & ... &  ... & 2 \\%No UV flux detected \\
124629 & \ 44.0233345 & \ 2.8086112 &  0.111 &     109 &     8 &   0.6564 &     109 &     8 &   0.1640 \\
124635 & \ 42.1595840 & 19.3280563 &  0.115 &     224 &     9 &   2.6102 &     224 &     9 &   0.7135 \\
171459 & 116.1820908 & 25.1405563 &  0.042 &     208 &    14 &   2.9989 &     208 &     9 &   0.4067 \\
174585 & 114.0429153 & \ 9.9863892 &  0.038 &    2540 &    30 &   2.0244 &    2531 &    30 &   0.5243 \\
174605 & 117.5900040 & \ 7.7944446 &  0.023 &     126 &    30 &   3.0337 &     126 &    30 &   0.6414 \\
182595 & 132.8004150 & 27.8800011 &  0.041 &    1689 &    20 &   2.1723 &    1689 &    20 &   0.5811 \\
191706 & 142.5533295 & 19.9905567 &  0.043 &     221 &    20 &   4.3573 &     221 &    18 &   0.6742 \\
191791 & 137.2241669 & 14.5838890 &  0.038 &    1626 &    15 &   0.7010 &    1626 &    12 &   0.1518 \\
\enddata
\tablecomments{Notes on sources: 1 = no GALEX data exists for this source;  2 = short exposure GALEX image exists, but source was not detected; 3 = no FUV GALEX image obtained for this source; FUV flux was derived using the measured NUV flux and Eq. 1.}
\tablecomments{Table 1 is published in its entirety in the machine-readable format.
      A portion is shown here for guidance regarding its form and content.}
\end{deluxetable*}

We highlight a few relevant points regarding the SHIELD galaxies by referring directly to Figure~\ref{fig:examples}.  First, visual inspection of the images readily reveals key features in these systems.  Regions of recent and current star formation are clearly evident in the NUV images of all four galaxies, coinciding with the locations of blue knots in the corresponding optical images.  For example, the bright NUV source at the southern end of AGC 111164 is co-spatial with the blue knot seen in the optical image, which in turn is seen to be a source of strong H$\alpha$ emission in our narrowband images \citep[see Figure 1 of][]{haurberg2015}.  These individual regions of star formation do not necessarily show up in the NIR images, where the light is dominated by older red giant stars.  

We also wish to reiterate the point made in Section~\ref{sec:shield} regarding the multiwavelength nature of the SHIELD dataset.  The combination of UV, optical, and NIR imaging for the full sample provides a powerful platform for studying the properties of these extremely low-mass galaxies, particularly when one considers the additional 21-cm \ion{H}{1} maps, narrowband H$\alpha$ images, and deep optical spectroscopy. 

\subsection{FUV SFRs for SHIELD Galaxies}\label{sec:sfr}
By determining flux measurements of SHIELD galaxies in the FUV images, we can determine the FUV star-formation rates for the SHIELD galaxies. The FUV fluxes were used to determine the FUV luminosities in units of ergs s$^{-1}$ Hz$^{-1}$ with the following equation:
\begin{equation}
  L_{FUV}=4\pi D^2 \times F_{FUV} \times 10^{0.4A_{FUV}},
\end{equation}
where D is the distance in cm, $F_{FUV}$ is the FUV flux density in ergs s$^{-1}$ cm$^{-2}$ Hz$^{-1}$, and $A_{FUV}$ is the FUV absorption in magnitudes.  Because the extinction along the line of sight in the FUV can be significant, we implement the following equation \citep{wyder2007} to determine the absorption at FUV wavelengths:
\begin{equation}
 A_{FUV}=8.24 \times E(B-V),
\end{equation}
where the color excess E(B - V) is used to represent the interstellar reddening. We note that this correction is for Galactic absorption only.  No corrections for absorption internal to the individual galaxies are applied in this study.  Given the extremely low-mass (and thus, low-metallicity) nature of the SHIELD sample we do not expect the internal absorption to be large for most of our sources.  Only a handful of SHIELD galaxies possess 24 micron Spitzer observations of the type often employed to derive absorption corrections to FUV fluxes \citep[e.g.,][]{Roychowdhury_2014}.  Based on our optical spectra for a subset of the SHIELD galaxies, we can use the Balmer decrement method to estimate their internal absorption, and find typical values of A$_{FUV}$ $\sim$ 0.3 mag, which result in a decrease in the inferred FUV SFRs of $\sim$0.12 dex.  We use these small estimated A$_{FUV}$ values to justify our decision to not apply an internal absorption correction to the full SHIELD sample.

Our UV fluxes are presented in Table \ref{tab:uvflux}.  We report our {\it measured} fluxes in Table~\ref{tab:uvflux} in units of counts s$^{-1}$, but convert them into cgs units for the derivation of the luminosities.  Column 1 gives the galaxy name (AGC number), columns 2 and 3 list the celestial coordinates for each source, and column 4 gives the E(B~-~ V) color excesses from \citep{schlegel1998} that are used to compute the Galactic absorption.  Columns 5 -- 7 list the total exposure time for each source for the relevant GALEX NUV image (in s), the apertures size used for measuring the source in the NUV image (in arcsec), and measured flux (in counts s$^{-1}$) obtained from the NUV measurement.   Columns 8 -- 10 present these same three values for the FUV measurements.  Finally, column 11 includes notes on individual sources, as specified in the Table notes.  We note in passing that Table~\ref{tab:uvflux} represents the first time that the {\it full sample} of 82 SHIELD galaxies are listed in print with their coordinates.  

The calculation of luminosities requires that the distance to each galaxy is known. The distances for the SHIELD galaxies were determined using one of two methods. The more precise method involves using TRGB distances from \citet{mcquinn2014} and \citet{mcquinn2021} when possible. The TRGB method uses the resolved stellar photometry from HST images to create a color-magnitude diagram (CMD) for each galaxy.  The CMDs are used to determine the apparent magnitudes of the TRGB which can then be used with the known TRGB absolute magnitude to determine distances. \citet{Savino_2022} have demonstrated that the TRGB provides a robust distance measurement for galaxies with M$_V$ $\le$ $-$9.5. The TRGB method is especially useful for dwarf galaxies, because the process has minimal metallicity dependence compared to other standard candles, and dwarf galaxies are relatively metal deficient. The second source of distances are flow-model distances. This process uses Hubble’s Law, corrected for any local gravitational effects, to calculate distance. This method is substantially less accurate than the TRGB method. Flow model distances for all ALFALFA detected galaxies are derived and included in the ALFALFA survey tables \citep{haynes2011, haynes2018}, based on the flow model of \citet{masters_2005}. 

FUV-based SFRs were calculated from our FUV luminosities using the conversion relation derived by  \citet{mcquinn2015b}:
\begin{equation}
 SFR_{FUV}=(2.04 \pm 0.81) \times 10^{-28} \cdot L_{FUV}, 
\end{equation}
where SFR$_{FUV}$ is in $M_\odot$ yr$^{-1}$ and L$_{FUV}$ is in ergs s$^{-1}$ Hz$^{-1}$.  We show in Section~\ref{sec:McQuinnSFRcomparison} that our measured FUV SFRs are quite robust, being in good agreement with SFR estimates derived using a completely independent method.   The distances as well as the FUV luminosities and SFRs for the SHIELD galaxies are given in Table \ref{tab:props}.

\subsection{Comparison with CMD-based SFRs}\label{sec:McQuinnSFRcomparison}

In this subsection we provide a comparison between the SHIELD FUV SFRs measured from GALEX fluxes with SFRs derived using CMD-based star-formation histories (SFHs) of 30 SHIELD galaxies \citep{mcquinn2015a, mcquinn2021}.  In brief, resolved stellar photometry obtained using relatively shallow HST images of the first 30 SHIELD galaxies was used to construct CMDs from which TRGB distances were determined \citep{mcquinn2014, mcquinn2021}.  These same CMDs were used to derive rough SFHs by fitting modeled CMDs created from synthetic stellar populations with a range of ages and metallicities and finding the best fit between the observed and modeled CMDs.  Among the derived parameters from this fitting method is the recent SFR of the galaxy, averaged over the last 200 Myr (SFR$_{200Myr}$).  Full details of the SFH modeling procedure are given in \citet{mcquinn2015a}.

\begin{figure}
\centering
\includegraphics[width=3.35in]{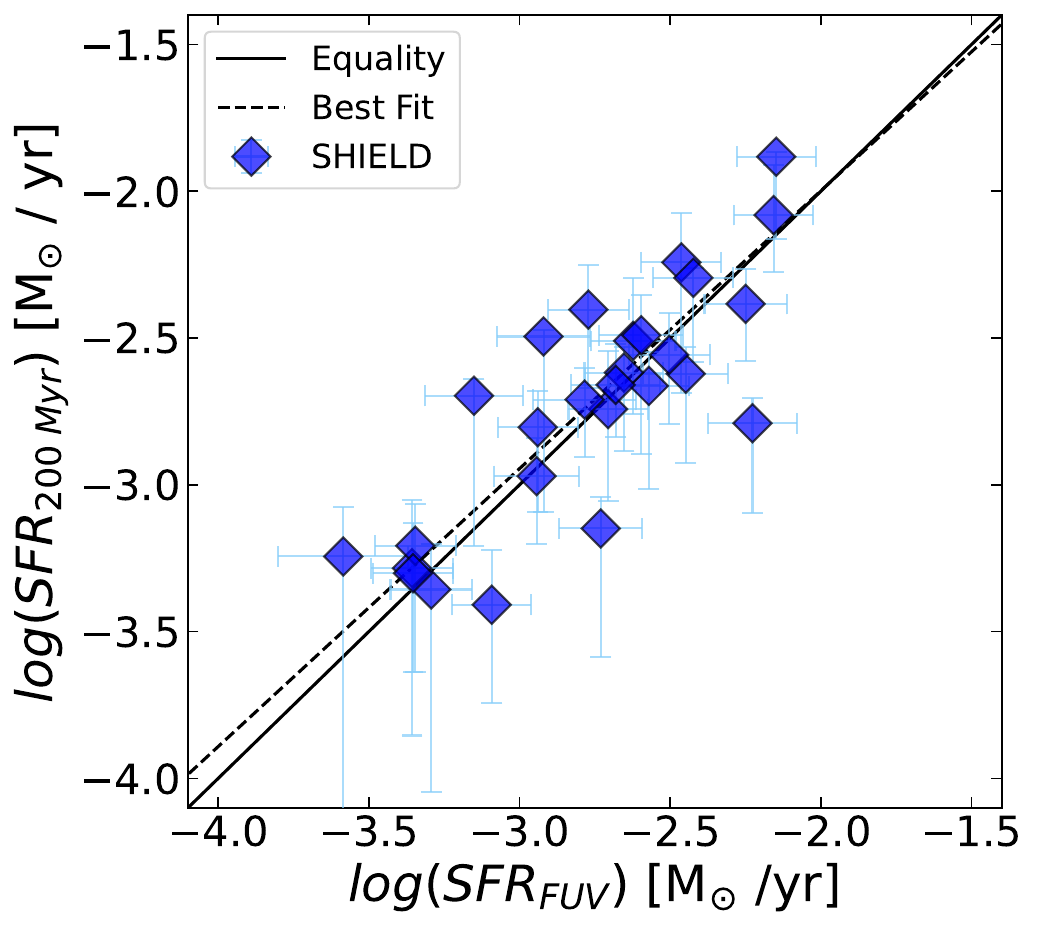}
\caption{Log-log graph of the CMD-based SFR$_{200Myr}$ (see text) vs. our FUV SFR. The solid line represents equality. The dashed line indicates a weighted bivariate best fit line with a slope of 0.95 ± 0.08. The orthogonal scatter about the best fit line is 0.21 dex.}
\label{fig:McQuinnSFRvsFUVSFR}
\end{figure}

We plot the SFR$_{200Myr}$ values tabulated in \citet{mcquinn2021} vs. the FUV SFRs derived in this paper in Figure~\ref{fig:McQuinnSFRvsFUVSFR}.  We expect reasonable agreement between these two quantities, since they are measuring the SFR over comparable timescales.  We find that the FUV SFRs correlate closely with the CMD-based 200 Myr SFRs, with a weighted bivariate best fit slope of 0.95 ± 0.08 and a weighted orthogonal scatter of 0.21 dex.  We consider the agreement between these two quantities excellent, given that they are derived using completely independent data and using dramatically different analysis methods.  This agreement is perhaps even more remarkable given that the quality of the shallow SHIELD CMDs used by \citet{mcquinn2014, mcquinn2021} is much lower than typically used for SFH determinations.

The general agreement between the two independent measures of the SFRs of the SHIELD galaxies shown in Figure~\ref{fig:McQuinnSFRvsFUVSFR} lends support to the hypothesis that our GALEX-based FUV SFRs are reliable.

%************************************************************

\section{Results: Star-Formation Rate Relationships for the SHIELD Galaxies}\label{sec:sfrrel}

We present the key results of this study using the FUV SFRs derived as described in the previous section.  For several of our results we compare and contrast the SFRs derived using both the FUV and H$\alpha$ methods.  The methodology for determining the H$\alpha$ SFRs is very similar to the process of deriving the FUV SFRs presented here.  The H$\alpha$ data and measurements are described in detail in Shepley et al. 2026 (in preparation).  Here we simply present the H$\alpha$ SFRs as needed for the current analysis.

Measured and derived quantities for the full sample of 82 SHIELD galaxies are presented in Table~\ref{tab:props}.  Column 1 lists the SHIELD galaxy name (AGC number), while column 2 lists the R-band apparent magnitude and its uncertainty from Shepley et al. 2026 (in preparation).  Column 3 presents the distances used for all of our distance-dependent quantities, while column 4 lists the sources for these distances.  As described in the previous section, these distances are a mix of TRGB distances and flow-model distances.  

%Column 5 lists the R-band absolute magnitude and its uncertainty, while columns 6 -- 8 give the stellar mass \citep{marine_2023}, \ion{H}{1} mass \citep{haynes2011, haynes2018}, and baryonic mass respectively (all in units of M$_\odot$).  Column 9 lists the FUV luminosity (in ergs s$^{-1}$ Hz$^{-1}$).  Columns 10 and 11 give FUV and H$\alpha$ SFRs in M$_\odot$ yr$^{-1}$. The values in Column 11 surrounded by parentheses are upper-limit H$\alpha$ SFRs (calculation described in Section~\ref{sec:sfrcomp}). Column 12 lists the FUV specific SFR (sSFR$_{FUV}$) (in yr$^{-1}$).  Finally, Column 13 presents the gas mass fraction M$_{gas}$/M$_{baryonic}$.  Notes describing the derivation of many of these quantities are presented below.

\startlongtable
\centerwidetable
\begin{longrotatetable}
\begin{deluxetable*}{cccccccccccccc}
\digitalasset
\tabletypesize{\scriptsize}

\tablecaption{Derived Properties of the SHIELD Galaxies\label{tab:props}}
\tablehead{
 AGC  \# & m$_R$  & Dist.  & Source & M$_R$  & log(M$_{star}$)  & log(M$_{HI}$)  & log(M$_{bary}$) &  log(L$_{FUV}$) & log(SFR$_{FUV}$)  & log(SFR$_{H\alpha}$) & log(sSFR$_{FUV}$)  & M$_{gas}$/M$_{bary}$ \\  %& GDT$_{FUV}$ \\
  & mag   & Mpc  &  & mag  & M$_\odot$  & M$_\odot$  & M$_\odot$  & erg/s/Hz &  M$_\odot$ yr$^{-1}$  &  M$_\odot$ yr$^{-1}$  & yr$^{-1}$  \\ %& & Gyr \\
(1) & (2) & (3) & (4) & (5) & (6) & (7) & (8) & (9) & (10) & (11) & (12) & (13) \\ %& (14)
} 
\startdata 
 102728 & 18.97 $\pm$ 0.08 & 12.41 $\pm$  0.64 & 1 & -11.60 $\pm$ 0.14 &  6.32 $\pm$ 0.20 &  7.05 $\pm$ 0.06 &  7.24 $\pm$ 0.06 & 24.75 $\pm$ 0.05 & -2.94 $\pm$ 0.14 & (-4.31 $\pm$ 0.16) &  -9.26 $\pm$ 0.24 & 0.88 $\pm$ 0.17 \\
 103722 & 17.25 $\pm$ 0.05 &  5.64 $\pm$  0.15 & 3 & -11.73 $\pm$ 0.07 &  6.43 $\pm$ 0.12 &  7.13 $\pm$ 0.03 &  7.32 $\pm$ 0.03 & 24.81 $\pm$ 0.06 & -2.89 $\pm$ 0.14 & -3.78 $\pm$ 0.06 &  -9.32 $\pm$ 0.19 & 0.87 $\pm$ 0.07 \\
 104208 & 19.55 $\pm$ 0.17 &  9.70 $\pm$  0.97 & 2 & -10.61 $\pm$ 0.28 &  5.99 $\pm$ 0.18 &  6.95 $\pm$ 0.10 &  7.11 $\pm$ 0.09 & 24.61 $\pm$ 0.14 & -3.08 $\pm$ 0.19 & (-4.47 $\pm$ 0.17) &  -9.07 $\pm$ 0.26 & 0.92 $\pm$ 0.28 \\
 110482 & 15.65 $\pm$ 0.03 &  7.82 $\pm$  0.21 & 1 & -14.03 $\pm$ 0.07 &  7.39 $\pm$ 0.12 &  7.28 $\pm$ 0.03 &  7.70 $\pm$ 0.06 & 25.23 $\pm$ 0.03 & -2.46 $\pm$ 0.13 & -2.87 $\pm$ 0.05 &  -9.85 $\pm$ 0.18 & 0.51 $\pm$ 0.08 \\
 111164 & 16.61 $\pm$ 0.03 &  5.11 $\pm$  0.07 & 1 & -12.06 $\pm$ 0.04 &  6.57 $\pm$ 0.12 &  6.60 $\pm$ 0.03 &  6.96 $\pm$ 0.05 & 24.60 $\pm$ 0.02 & -3.09 $\pm$ 0.13 & -3.48 $\pm$ 0.05 &  -9.66 $\pm$ 0.18 & 0.59 $\pm$ 0.08 \\
 111946 & 17.26 $\pm$ 0.03 &  9.02 $\pm$  0.29 & 1 & -12.70 $\pm$ 0.08 &  6.68 $\pm$ 0.12 &  7.16 $\pm$ 0.03 &  7.39 $\pm$ 0.04 & 25.12 $\pm$ 0.03 & -2.57 $\pm$ 0.13 & -3.39 $\pm$ 0.06 &  -9.25 $\pm$ 0.18 & 0.80 $\pm$ 0.09 \\
 111977 & 15.43 $\pm$ 0.03 &  5.96 $\pm$  0.11 & 1 & -13.61 $\pm$ 0.05 &  7.10 $\pm$ 0.12 &  6.85 $\pm$ 0.03 &  7.35 $\pm$ 0.07 & 24.98 $\pm$ 0.02 & -2.71 $\pm$ 0.13 & -3.27 $\pm$ 0.05 &  -9.81 $\pm$ 0.18 & 0.43 $\pm$ 0.08 \\
 112503 & 16.73 $\pm$ 0.03 & 10.20 $\pm$  1.02 & 2 & -13.44 $\pm$ 0.22 &  7.31 $\pm$ 0.17 &  7.14 $\pm$ 0.10 &  7.59 $\pm$ 0.10 & ... $\pm$ ... & ... $\pm$ ... & -2.60 $\pm$ 0.10 &  ... $\pm$ ... & 0.48 $\pm$ 0.15 \\
 112505 & 20.68 $\pm$ 0.34 & 10.30 $\pm$  1.03 & 2 &  -9.53 $\pm$ 0.41 &  6.10 $\pm$ 0.17 &  7.12 $\pm$ 0.09 &  7.28 $\pm$ 0.09 & 24.71 $\pm$ 0.09 & -2.98 $\pm$ 0.16 & (-4.46 $\pm$ 0.17) &  -9.08 $\pm$ 0.23 & 0.93 $\pm$ 0.27 \\
 112521 & 17.62 $\pm$ 0.03 &  6.58 $\pm$  0.18 & 1 & -11.61 $\pm$ 0.07 &  6.25 $\pm$ 0.12 &  6.85 $\pm$ 0.03 &  7.05 $\pm$ 0.03 & 24.33 $\pm$ 0.03 & -3.36 $\pm$ 0.13 & -4.09 $\pm$ 0.07 &  -9.61 $\pm$ 0.18 & 0.84 $\pm$ 0.10 \\
\\
 123352 & 18.65 $\pm$ 0.08 &  9.72 $\pm$  0.25 & 1 & -11.85 $\pm$ 0.09 &  6.57 $\pm$ 0.12 &  7.18 $\pm$ 0.03 &  7.38 $\pm$ 0.03 & 25.46 $\pm$ 0.07 & -2.23 $\pm$ 0.15 & -3.03 $\pm$ 0.06 &  -8.80 $\pm$ 0.19 & 0.85 $\pm$ 0.08 \\
 124056 & 20.82 $\pm$ 0.29 &  5.90 $\pm$  0.59 & 2 &  -8.40 $\pm$ 0.37 &  6.42 $\pm$ 0.17 &  6.55 $\pm$ 0.10 &  6.87 $\pm$ 0.09 & ... $\pm$ ... & ... $\pm$ ... & (-4.84 $\pm$ 0.17) &  ... $\pm$ ... & 0.64 $\pm$ 0.19 \\
 124629 & 19.53 $\pm$ 0.20 & 10.60 $\pm$  1.06 & 2 & -10.85 $\pm$ 0.30 &  6.21 $\pm$ 0.17 &  7.06 $\pm$ 0.10 &  7.23 $\pm$ 0.09 & 24.74 $\pm$ 0.13 & -2.95 $\pm$ 0.19 & (-4.39 $\pm$ 0.17) &  -9.16 $\pm$ 0.25 & 0.90 $\pm$ 0.27 \\
 124635 & 18.00 $\pm$ 0.05 &  7.90 $\pm$  0.79 & 2 & -11.75 $\pm$ 0.22 &  7.19 $\pm$ 0.17 &  7.05 $\pm$ 0.09 &  7.49 $\pm$ 0.10 & 25.14 $\pm$ 0.09 & -2.55 $\pm$ 0.16 & -2.57 $\pm$ 0.10 &  -9.74 $\pm$ 0.23 & 0.50 $\pm$ 0.15 \\
 171459 & 17.29 $\pm$ 0.03 & 11.80 $\pm$  1.18 & 2 & -13.17 $\pm$ 0.22 &  6.94 $\pm$ 0.17 &  7.21 $\pm$ 0.09 &  7.48 $\pm$ 0.08 & 25.00 $\pm$ 0.10 & -2.69 $\pm$ 0.16 & -3.14 $\pm$ 0.10 &  -9.63 $\pm$ 0.24 & 0.71 $\pm$ 0.21 \\
 174585 & 17.38 $\pm$ 0.04 &  7.89 $\pm$  0.21 & 1 & -12.19 $\pm$ 0.07 &  6.56 $\pm$ 0.12 &  6.90 $\pm$ 0.04 &  7.16 $\pm$ 0.04 & 24.75 $\pm$ 0.03 & -2.94 $\pm$ 0.13 & -3.28 $\pm$ 0.05 &  -9.50 $\pm$ 0.18 & 0.75 $\pm$ 0.10 \\
 174605 & 17.17 $\pm$ 0.04 & 10.89 $\pm$  0.28 & 1 & -13.07 $\pm$ 0.07 &  6.94 $\pm$ 0.12 &  7.27 $\pm$ 0.03 &  7.53 $\pm$ 0.04 & 25.07 $\pm$ 0.05 & -2.62 $\pm$ 0.14 & -3.31 $\pm$ 0.05 &  -9.56 $\pm$ 0.18 & 0.74 $\pm$ 0.09 \\
 182595 & 16.17 $\pm$ 0.04 &  9.02 $\pm$  0.28 & 1 & -13.70 $\pm$ 0.08 &  7.22 $\pm$ 0.12 &  6.91 $\pm$ 0.04 &  7.44 $\pm$ 0.07 & 24.92 $\pm$ 0.03 & -2.77 $\pm$ 0.13 & -2.72 $\pm$ 0.05 &  -9.99 $\pm$ 0.18 & 0.40 $\pm$ 0.08 \\
 191706 & 16.67 $\pm$ 0.03 &  8.20 $\pm$  0.82 & 2 & -12.99 $\pm$ 0.22 &  7.08 $\pm$ 0.17 &  7.20 $\pm$ 0.09 &  7.52 $\pm$ 0.08 & 24.91 $\pm$ 0.09 & -2.78 $\pm$ 0.16 & (-4.68 $\pm$ 0.17) &  -9.86 $\pm$ 0.23 & 0.64 $\pm$ 0.18 \\
 191791 & 17.34 $\pm$ 0.02 &  9.50 $\pm$  0.95 & 2 & -12.64 $\pm$ 0.22 &  6.77 $\pm$ 0.17 &  6.78 $\pm$ 0.16 &  7.14 $\pm$ 0.12 & 24.37 $\pm$ 0.09 & -3.32 $\pm$ 0.16 & (-4.55 $\pm$ 0.17) & -10.09 $\pm$ 0.23 & 0.58 $\pm$ 0.27 \\
\enddata
\tablecomments{Source of distances (Column 4): 1 = distance based on TRGB method, taken from \citet{mcquinn2014} and \citet{mcquinn2021}; 2 = distance based on flow model of \citet{masters_2005}, taken from AFLALFA catalogs \citep{haynes2011, haynes2018}; 3 = distance derived using TRGB method, based on unpublished analysis (K. McQuinn, private communication).}
\tablecomments{H$\alpha$ SFRs (Column 11) enclosed in parentheses represent upper limits.}
\tablecomments{Table 2 is published in its entirety in the machine-readable format.
      A portion is shown here for guidance regarding its form and content.}
\end{deluxetable*}
\end{longrotatetable}

%************************************************************

%Measured and derived quantities for the full sample of 82 SHIELD galaxies are presented in Table~\ref{tab:props}.  Column 1 lists the SHIELD galaxy name (AGC number), while column 2 lists the R-band apparent magnitude and its uncertainty from Shepley et al. 2026 (in preparation).  Column 3 presents the distances used for all of our distance-dependent quantities, while column 4 lists the sources for these distances.  As described in the previous section, these distances are a mix of TRGB distances and flow-model distances.  
Column 5 lists the R-band absolute magnitude and its uncertainty, while columns 6 -- 8 give the stellar mass \citep{marine_2023}, \ion{H}{1} mass \citep{haynes2011, haynes2018}, and baryonic mass respectively (all in units of M$_\odot$).  Column 9 lists the FUV luminosity (in ergs s$^{-1}$ Hz$^{-1}$).  Columns 10 and 11 give FUV and H$\alpha$ SFRs in M$_\odot$ yr$^{-1}$. The values in Column 11 surrounded by parentheses are upper-limit H$\alpha$ SFRs (calculation described in Section~\ref{sec:sfrcomp}). Column 12 lists the FUV specific SFR (sSFR$_{FUV}$) (in yr$^{-1}$).  Finally, Column 13 presents the gas mass fraction M$_{gas}$/M$_{baryonic}$.  Notes describing the derivation of many of these quantities are presented below.

\subsection{The SHIELD Sample in Context: Description of Comparison Data Sets}\label{sec:compare}

Here we describe the various comparison data sets we use to place our results for the SHIELD galaxies into a broader context.
The selection of our particular comparison samples exactly mimics the choices made in \citet{teich2016}.  We do this to provide continuity between the previous and current SHIELD-based studies.  Our comparison data sets are drawn from: 11HUGS \citep{Lee_2009}, THINGS \citep{Walter_2008}, LITTLE-THINGS \citep{hunter2012}, FIGGS \citep{Roychowdhury_2014}, and VLA-ANGST \citep{ott2012}. The latter 3 samples consist primarily of nearby dwarf galaxies, while 11HUGS serves as a comprehensive sample of galaxies in the local universe (distances out to 11 Mpc) and THINGS serves as a comparison of \ion{H}{1}-detected galaxies in the local universe (distances of 2-15 Mpc).  

\subsubsection{11HUGS}\label{11HUGS}
The data for the 11HUGS galaxies were primarily taken from \citet{Lee_2009}. The stellar masses were obtained from \citet{Dale_2023}, who published a catalog of M$_{stellar}$ values for the Local Volume Legacy (LVL) sample which has significant overlap with the 11HUGS galaxies.  This resulted in 115 stellar masses.  For the \ion{H}{1} masses, 65 masses were obtained from the ALFALFA survey \citep{haynes2011, haynes2018}, and 37 masses were obtained from galaxies that overlapped with the VLA-ANGST, FIGGS and LITTLE-THINGS, totaling 102 \ion{H}{1} masses for 11HUGS.

\subsubsection{THINGS}\label{THINGS}
The THINGS sample is used to provide a higher-mass (i.e., non-dwarf) comparison sample of galaxies with measured M$_{HI}$ masses.  The THINGS galaxies appear only in Figure \ref{fig:HImasscomparisonplot} (plot of $\log(M_{HI})$ vs. $\log(SFR_{FUV})$), to maintain consistency between our plot and the similar plot from \citet{teich2016}. The THINGS data were obtained from \citet{Walter_2008}.

\subsubsection{LITTLE-THINGS}\label{LT}
The majority of the LITTLE THINGS data were obtained from \citet{hunter2012}. The SFRs published in \citet{hunter2012} are normalized to the surface area of the disks of the galaxies (units of M$_\odot$ yr$^{-1}$ kpc$^{-2}$), so the SFRs used here were instead obtained from \citet{Hunter_2010}. It should be noted that this creates a discrepancy between the LITTLE-THINGS SFRs used here and those used in \citet{teich2016}.  Overall, the changes are not large since the disk scale lengths of the LITTLE THINGS galaxies are of order 1 kpc, meaning that the conversion from SFR to disk-area-normalized SFR is typically of order unity for this sample.  The absolute magnitudes presented in \citet{hunter2012} are measured in the V-band, which we have converted to the B-band using the average B-V value of the FIGGS sample in order to maintain consistency across all comparison samples.

\subsubsection{FIGGS}\label{FIGGS}
The FIGGS data were obtained from \citet{begum2008} and \citet{Roychowdhury_2014}.  The stellar masses for the FIGGS galaxies were calculated using standard mass-to-light ratio models and the Bell \& de Jong relation \citep{Bell_2001}. The SFRs were calculated from the H$\alpha$ and FUV luminosities given in \citet{Roychowdhury_2014} using the same conversion factors used for the SHIELD galaxies  \citep[e.g.,][]{Kennicutt1998a, mcquinn2015b}. This maintains consistency between the FIGGS data and the SHIELD data. This conversion was {\it not} used for all of the comparison samples because the suggested SFR range in \citet{mcquinn2015b} to which the FUV conversion applies is $10^{-3}-10^{-1}$ $M_\odot/yr$, and the other samples have a number of galaxies that fall outside of this range.

\subsubsection{VLA-ANGST}\label{VLA-ANGST}
Nearly all of the VLA-ANGST data were obtained from \citet{ott2012}. Stellar masses for the VLA-ANGST sample have not been formally published, so the stellar masses were obtained from the other comparison samples when there were overlapping galaxies. This resulted in stellar masses for 19 galaxies. It should be noted that no H$\alpha$ data has been acquired for the VLA-ANGST sample, so these galaxies are not present in plots that contain H$\alpha$ SFRs.

\begin{figure*}
\centering
\includegraphics[width=6.0in]{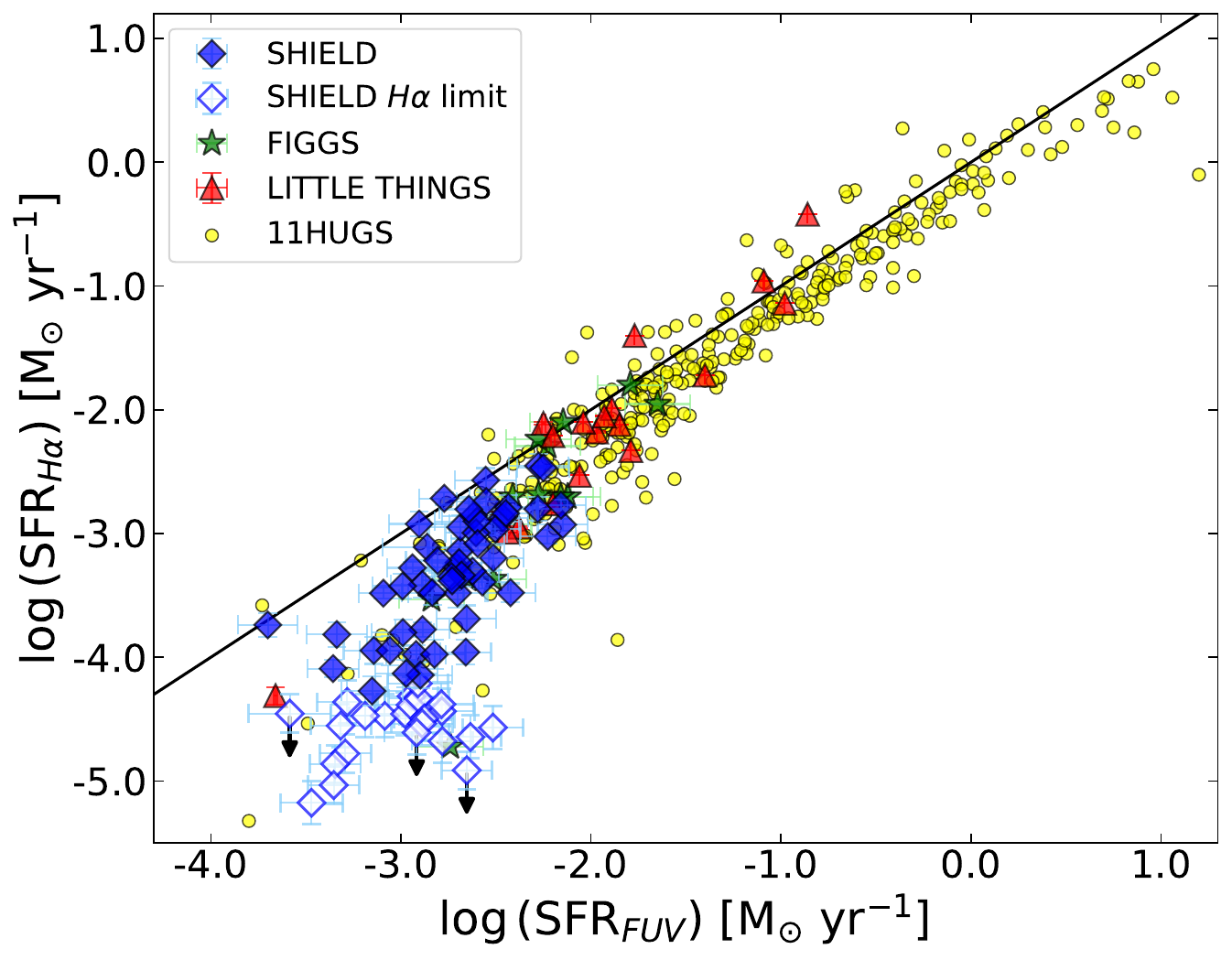}
\caption{ Log-log graph of H$\alpha$ SFR vs.\ FUV SFR, in units of solar masses per year, for SHIELD as well as other nearby galaxy surveys. The open diamonds indicate SHIELD galaxies whose H$\alpha$ fluxes represent 3$\sigma$ upper limits; upper-limit arrows are shown on only a few of these open diamonds. The solid line represents equality.}
\label{fig:SFRcomparisonplot1}
\end{figure*}

\begin{figure*}
\centering
\includegraphics[width=6.0in]{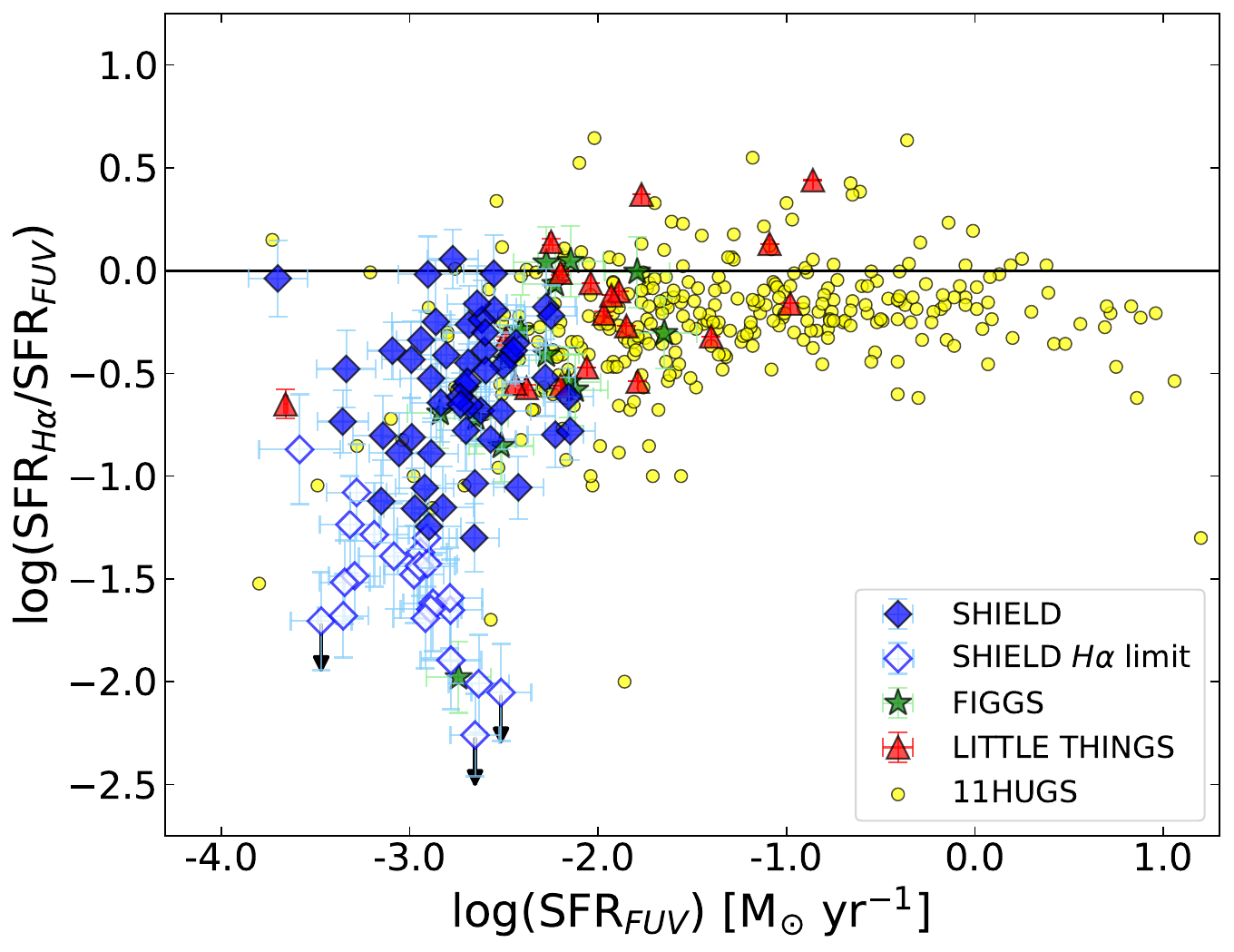}
\caption{Log-log graph of (H$\alpha$ SFR/FUV SFR) vs.\ FUV SFR for SHIELD as well as other nearby galaxy surveys. The open diamonds indicate SHIELD galaxies whose H$\alpha$ fluxes represent 3$\sigma$ upper limits; upper-limit arrows are shown on only a few of these open diamonds.  The solid line indicates $(SFR_{H\alpha}/SFR_{FUV})=1$.}
\label{fig:SFRcomparisonplot2}
\end{figure*}

\subsection{H$\alpha$ SFR vs.\ FUV SFR}\label{sec:sfrcomp}
We begin by comparing the FUV SFRs to the H$\alpha$ SFRs.  Not all of the SHIELD galaxies had measurable H$\alpha$ flux present in their narrowband images.  Only 55 of 82 (67.1\%) galaxies had detectable H$\alpha$ flux, while the remaining 27 galaxies were not detected.  In cases where the galaxy was undetectable, an upper limit flux was used for the galaxies.  This value was determined by taking the median flux uncertainty for the ten {\it detected} SHIELD galaxies with the lowest measured fluxes and multiplying by 3.0.  This provides a conservative 3$\sigma$ upper limit on the H$\alpha$ flux for our undetected sources, which was then used to calculate an upper limit for the H$\alpha$ luminosity and SFR.  Full details will be presented in Shepley et al. 2026 (in preparation).

In Figure~\ref{fig:SFRcomparisonplot1}, we plot $\log(SFR_{H\alpha})$ vs.\ $\log(SFR_{FUV})$ for the SHIELD galaxies.   This figure mirrors Figure 16 in \citet{teich2016}, with the primary difference being that Figure~\ref{fig:SFRcomparisonplot1} includes the full SHIELD sample while the \citet{teich2016} plot shows only the original 12 SHIELD galaxies.   Similar to \citet{teich2016} we also plot these same quantities for galaxies from three comparison samples: 11HUGS \citep{Lee_2009}, LITTLE-THINGS \citep{hunter2012}, and FIGGS \citep{Roychowdhury_2014}.  For the SHIELD sample, the open diamonds indicate galaxies for which the H$\alpha$ flux is given by a 3$\sigma$ upper limit. 

The full SHIELD sample dramatically fills in the lower SFR regions (log(SFR$_{FUV}$) $<$ $-$2.0 M$_\odot$ yr$^{-1}$) of Figure~\ref{fig:SFRcomparisonplot1} as compared to the equivalent plot from the \citet{teich2016} study.   The SHIELD sources overlap galaxies from both of the two dwarf comparison samples (LITTLE THINGS and FIGGS), although the SHIELD systems are on average probing to lower SFRs.  For example, the median SFR values for both $H\alpha$ and FUV for the LITTLE THINGS sample lie above the {\it highest} SFRs for the SHIELD galaxies.

The new SHIELD SFRs corroborate and extend the well-known result that the H$\alpha$ SFRs of dwarf galaxies tend to be systematically lower than their FUV SFRs \citep[e.g.,][]{sullivan2000,bell&kennicutt2001,salim2007,Lee_2009,Roychowdhury_2014}.  As seen in Figure~\ref{fig:SFRcomparisonplot1}, the departure from SFR$_{H\alpha}$ $\approx$ SFR$_{FUV}$ starts near log(SFR$_{FUV}$) = $-$2.0, very close to the location where the SHIELD galaxies start to appear in the plot.  The deviation between the two SFR measures increases strongly as lower SFR values are reached.  This deviation is generally attributed to the effect of stochasticity in the population of the high-mass end of the IMF in the low-mass, low-SFR regime, combined with the non-uniform sampling of the star-formation history when the star formation has proceeded via a series of distinct events \citep{fumagalli2011,roychowdhury2011,dasilva2012,weisz2012}.  Very nearly all of the SHIELD galaxies are located in this regime, and show large departures from SFR$_{H\alpha}$ $\approx$ SFR$_{FUV}$.

\begin{figure*}
\centering
\includegraphics[width=6.0in]{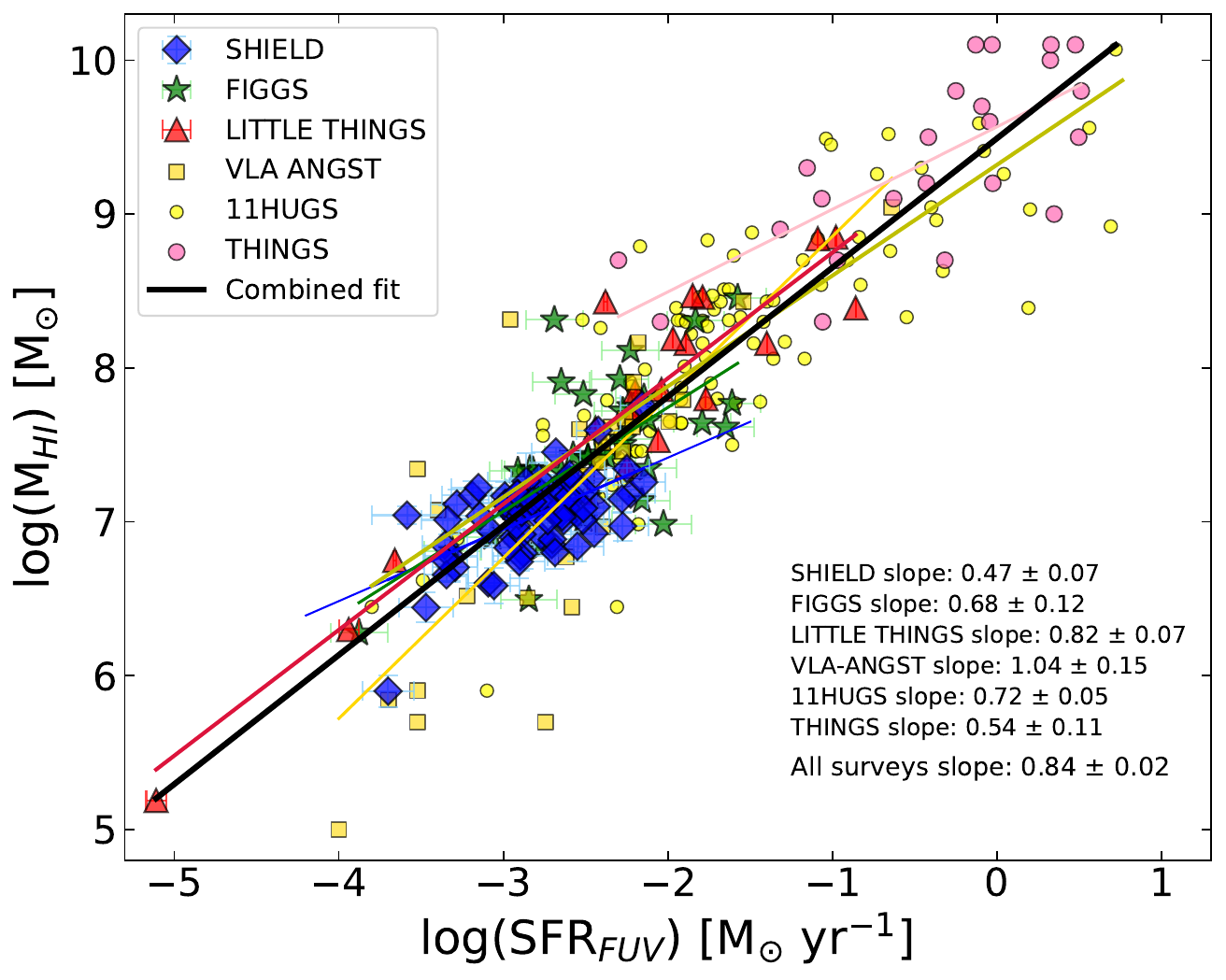}
\caption{Log-log graph of \ion{H}{1} gas mass vs.\ FUV SFR for SHIELD galaxies as well as other nearby galaxy surveys. The slope of the best fit and its uncertainty has been calculated for each sample, as well as a best fit line and its uncertainty for all of the samples combined. The SHIELD best fit line has been extended for clarity.}
\label{fig:HImasscomparisonplot}
\end{figure*}

The departure of the two SFRs from equality is more clearly illustrated in Figure~\ref{fig:SFRcomparisonplot2}, which plots the logarithm of the ratio of H$\alpha$ to the FUV SFRs vs. log(SFR$_{FUV}$).  The SHIELD galaxies follow a strong downward trend which starts around log(SFR$_{FUV}$) = $-$2. Nearly all of the galaxies shown in the plot with log(SFR$_{FUV}$) $<$ $-$2.5 are less clustered around the unity line compared to the galaxies with higher SFRs.  The most extreme cases -- mostly H$\alpha$ upper limits -- have SFR$_{H\alpha}$ two orders of magnitude smaller than SFR$_{FUV}$.

\subsection{Mass Quantities vs. FUV SFR}\label{sec:Masses}

In this subsection, we compare the FUV star-formation rates of the SHIELD galaxies with various mass or mass-like quantities.  Specifically, we consider how the SFR scales with \ion{H}{1} mass, stellar mass, baryonic mass, and B-band absolute magnitude.  After presenting each of the various empirical relations in turn, we inter-compare them in Section~\ref{sec:masssummary}.

We note here that all of the fits presented in this section are {\it unweighted} linear least-squares fits, due to the fact that uncertainties are not consistently available for all of the quantities being presented for the comparison samples.

\subsubsection{\ion{H}{1} Mass vs. FUV SFR}\label{sec:HI}
Following \citet[][see their Figure 14]{teich2016}, we first compare the FUV star formation rates of the SHIELD galaxies to their \ion{H}{1} masses.  Our presentation includes the same set of comparison galaxies used by \citet{teich2016}, but with the addition of 11HUGS.  This results in a total sample that covers a factor of $\sim$10$^5$ in \ion{H}{1} mass and a factor of $\sim$10$^6$ in FUV SFR.  %This comparison aims to improve our understanding of the process by which dwarf galaxies convert their neutral hydrogen to stars. 
The \ion{H}{1} masses for the SHIELD galaxies are computed using the \ion{H}{1} fluxes cataloged in the ALFALFA survey paper \citep{haynes2011, haynes2018}, but using our updated TRGB distances when these are available.

In Figure~\ref{fig:HImasscomparisonplot}, we plot $\log(SFR_{FUV})$ vs.\ $\log(M_{HI})$ for SHIELD galaxies, as well as for the 11HUGS \citep{Lee_2009}, LITTLE-THINGS \citep{hunter2012}, THINGS \citep{Walter_2008}, FIGGS \citep{Roychowdhury_2014} and VLA-ANGST \citep{ott2012} surveys.  Following \citet{teich2016}, we fit a linear regression to the data from each survey independently, as well as a combined fit which includes all galaxies from all surveys presented.  The galaxies from the six different surveys are plotted with different symbols, as shown in the legend in the upper left corner of the plot, while the slopes of the various linear fits are listed in the lower right corner.  The best-fit lines are shown in the same color as the data points from that sample (e.g., the pink line is the fit to the THINGS galaxies), and the lengths of each of the best-fit lines are limited to the range of the FUV SFRs for each sample (with the exception of the SHIELD galaxy fit, which is extended for improved visibility).

The simple expectation is that there should be a correlation between \ion{H}{1} mass in a galaxy and its SFR.  However, this correlation would not necessarily be expected to be a tight one, since stars form from colder, denser molecular gas rather than directly from \ion{H}{1} gas.   Both of these expectations are borne out in the data shown in Figure~\ref{fig:HImasscomparisonplot}.  While there is indeed a strong correlation between the quantities being plotted, the scatter about the best-fit lines is substantial.  In particular, the VLA ANGST sample shows a very large scatter in Figure~\ref{fig:HImasscomparisonplot}.  Interestingly, the SHIELD galaxies exhibit a fairly tight relation between \ion{H}{1} mass and FUV SFR.  However, these SHIELD galaxies also have the most shallow slope of any of the six galaxy samples.  This is presumably due to the way the SHIELD galaxies were selected, which results in a very narrow \ion{H}{1} mass range which covers only a factor of $\sim$ten.

%*******************************************************************************

\subsubsection{Stellar Mass vs.\ FUV SFR}\label{sec:stellarmass}

\begin{figure}
\centering
\includegraphics[width=3.35in]{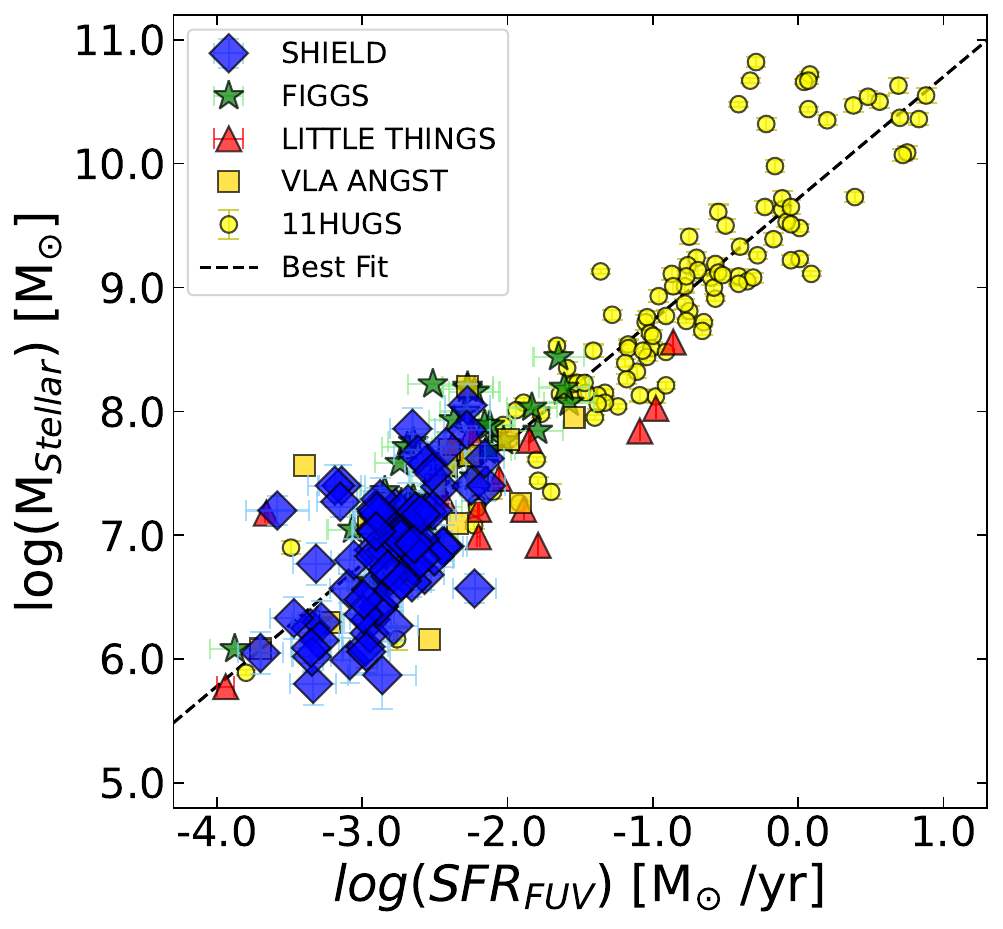}
\caption{Log-log graph of Stellar Mass vs.\ FUV SFR. The dashed line indicates a best fit line with a slope of 0.98 ± 0.02. The scatter about the best fit line is 0.42 dex.}
\label{fig:stellarmassvssfr}
\end{figure}

Next we compare the stellar masses of the SHIELD galaxies to their FUV SFRs. 
%for SHIELD galaxies and our comparison samples to gain a better understanding of how stellar mass correlates to FUV SFR. 
The results are shown in Figure~\ref{fig:stellarmassvssfr}, where we plot $\log(M_{Stellar})$ vs.\ $\log(SFR_{FUV})$.  We again include data from our various comparison samples, as indicated in the legend in the upper left.   For the SHIELD galaxies the stellar masses come from \citet{marine_2023}, who used Spitzer images obtained at 3.6 $\mu$m to derive NIR luminosities for all 82 SHIELD sources.   The masses were computed by adopting a stellar mass-to-light ratio at 3.6 $\mu$m of 0.50 \citep{mcgaugh2014, lelli2016}.

We see that there is a strong correlation between stellar mass and FUV SFR.  The dashed line shows the best-fit linear relation which has a slope of 0.98 ± 0.02.  
In Figure~\ref{fig:stellarmassvssfr}, and all subsequent figures in  this subsection, we plot only the best-fit linear relation for the entire sample of galaxies included in the graph. The SHIELD galaxies strongly define the observed relationship at the low-mass end, and generally follow the same trend that is defined by the other low-mass galaxies from the comparison samples.  

%*******************************************************************************

\subsubsection{Baryonic Mass vs.\ FUV SFR}\label{sec:baryonicmass}
\begin{figure}
\centering
\includegraphics[width=3.35in]{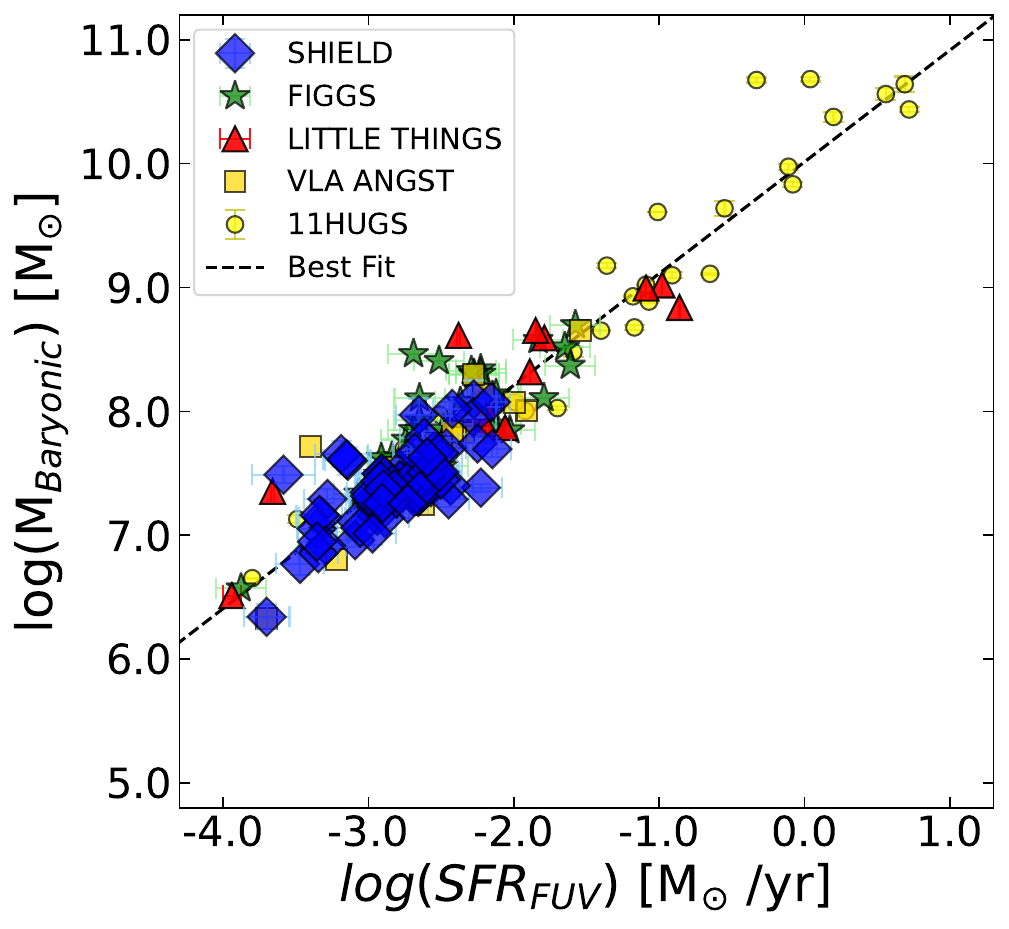}
\caption{Log-log graph of Baryonic Mass vs.\ FUV SFR. The dashed line indicates a best fit line with a slope of 0.90 ± 0.02. The scatter about the best fit line is 0.26 dex. }
\label{fig:BaryonicMassPlot}
\end{figure}

We compare the baryonic mass to the FUV SFR for the SHIELD galaxies as well as our comparison samples 
in Figure~\ref{fig:BaryonicMassPlot}.  The baryonic mass is defined as 
\begin{equation}
  M_{baryonic} = M_{stellar} + M_{gas},
\end{equation}
where
\begin{equation}
  M_{gas} = 1.35 \times M_{HI}.
\end{equation}
Here the factor of 1.35 accounts for the mass fraction of helium gas in the ISM of the galaxies.  No effort is made to account for any molecular gas present in these systems, since the direct measurement of molecular gas via the detection of the CO rotational lines is notoriously  difficult in metal-poor galaxies \citep{sage1992, Taylor_1998, leroy2005, schruba2012}.

Once again we see that the SHIELD galaxies follow the same trend as the other dwarf samples. The relationship between $\log(M_{baryonic})$ vs.\ $\log(SFR_{FUV})$ for the SHIELD galaxies shown in Figure~\ref{fig:BaryonicMassPlot} exhibits a lower scatter (0.26 dex) than the scatter present in the previous plots.  This point will be discussed in more detail in Section~\ref{sec:masssummary}.  The dashed best-fit line has a slope of 0.90 ± 0.02.
%, indicating that a galaxy's baryonic mass and FUV SFR correlate to the first power. 
%This correlation will be compared to the correlation between other mass quantities vs FUV SFR in Section~\ref{sec:Discussion}.

\subsubsection{Absolute Magnitude vs.\ FUV SFR}\label{sec:absmag}
A galaxy's absolute magnitude is often seen as an analog of the stellar mass of the galaxy, particularly when measured at longer wavelengths. In this subsection we compare the B-band absolute magnitudes to the FUV SFR of the SHIELD galaxies.  
%to better understand how a galaxy’s absolute magnitude affects its SFR on different time scales. 
We chose to use M$_B$ for this comparison since the majority of the comparison samples provide B-band values.  The absolute magnitudes of the SHIELD galaxies were converted from the R-band to the B-band using the conversion given in \citet{salzer_1989}. %The results are shown in Figure~\ref{fig:absmagplot}.

\begin{figure}
\centering
\includegraphics[width=3.35in]{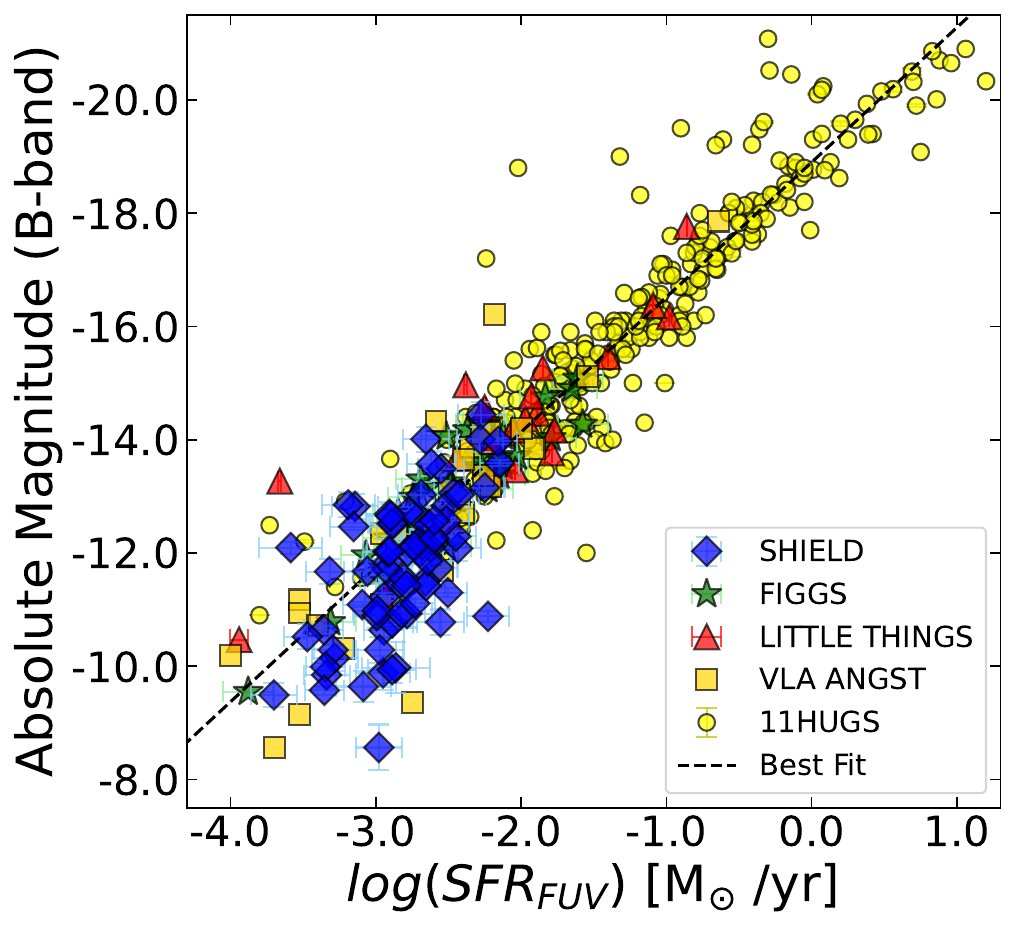}
\caption{Graph of the B-band absolute magnitude vs.\ the log of FUV SFR. The dashed line indicates a best fit line with a slope of 2.38 ± 0.04. The scatter about the best fit line is 0.87 magnitudes.}
\label{fig:absmagplot}
\end{figure}

In Figure~\ref{fig:absmagplot}, we plot B-band absolute magnitude (M$_B$) vs.\ $\log(SFR_{FUV})$.  The trend across all of the galaxy surveys reveals that a brighter absolute magnitude (higher stellar mass) is associated with a higher FUV SFR. This is expected given that absolute magnitude scales with stellar mass in a predictable way.  
%for star-forming galaxies, because the higher absolute magnitude is a result of a larger stellar mass in the galaxy, and the stellar mass increases with a galaxy’s SFR.  
We include the same comparison samples employed in the previous plots in this subsection.
%, although we have opted to include only those galaxies from the dwarf-dominated studies in Figure~\ref{fig:absmagplot}.  
Visual inspection suggests a larger scatter in this ``mass" - SFR comparison, particularly at the low-luminosity end.

The best-fit linear relation shown in Figure~\ref{fig:absmagplot} has a slope of 2.38 $\pm$ 0.04 and an RMS scatter of 0.87 magnitudes.\footnote{In order to better compare these values with those from the other plots in this subsection, one could imagine replacing the absolute magnitudes with the logarithm of the B-band luminosity (L$_B$).  In this case, the slope and scatter would be reduced by a factor of 2.5, resulting in a slope of 0.95 and an RMS scatter of 0.35 dex.}  The large scatter seen in the figure, particularly for the lower luminosity systems, is perhaps not surprising.  Galaxies at a fixed FUV SFR might be expected to exhibit a range of luminosity variation due to the difference in the timescales for the enhancement of the B-band light compared with the FUV.  The variations for the lower masses would be enhanced simply because the relative contribution to the total light from the galaxy by the younger stars would be large, giving rise to the substanial scatter observed.  For the more massive galaxies in the upper half of the diagram, the corresponding impact on the B-band luminosity would be less dramatic, even though the SFRs are larger, since the relative contribution to the galaxy light by the younger stars would be smaller.

\subsubsection{Summary of ``Mass" vs.\ FUV SFR Relations}\label{sec:masssummary}

Table~\ref{tab:masscomparisons} summarizes the results of our four ``mass" vs. FUV SFR relations presented in this subsection of the paper.  One of the goals of carrying out these different mass -- FUV SFR comparisons was to determine which of these mass quantities is the best tracer of the SFR in our galaxies.  In addition to the best-fit lines computed for the full samples of galaxies, we also carried out fits to just the dwarf galaxies (SHIELD, LITTLE THINGS, FIGGS, and VLA ANGST) and to the SHIELD galaxies alone.  The results of these fits are all included in Table~\ref{tab:masscomparisons}.  Note that even though we plot (and fit to) M$_B$ in Figure~\ref{fig:absmagplot}, we present the slopes and scatters for the fits in Table~\ref{tab:masscomparisons} as if we were using log(L$_B$) instead.  This allows for a more direct comparison with the results from the three mass plots.  Note also that we have opted to not include the THINGS sample in the ``All Surveys" fit for the \ion{H}{1} Mass -- FUV SFR relation presented in the table.  This means that we are using the same comparison samples for all four fits, but it also means that the slopes listed in Figure~\ref{fig:HImasscomparisonplot} and Table~\ref{tab:masscomparisons} for ``All Surveys" are slightly different.

Examination of Table~\ref{tab:masscomparisons} shows that the best fit slopes to the ``All Surveys" data samples all have values close to unity, indicating a direct relationship between mass and SFR (i.e., mass $\propto$ FUV SFR).  This is particularly true for the stellar mass, baryonic mass, and absolute magnitude fits.  Direct scaling between mass and SFR is expected for these gas-rich galaxy samples, particularly when considering the FUV SFR since the latter should be driven by the total mass of OBA stars created over the past few hundred million years.  It is interesting to note that the slope presented in column 2 of Table~\ref{tab:masscomparisons} that is farthest from 1.0 is the \ion{H}{1} Mass -- FUV SFR relation.  This is perhaps no surprise since, as mentioned above, stars are thought to form from molecular gas rather than neutral hydrogen gas.  

\begin{deluxetable*}{ccccccccccc}
\tabletypesize{\scriptsize}
%\tablewidth{0pt} 
%\tablenum{1}
\tablecaption{Results of Linear Regressions for Mass vs.\ FUV SFR Comparisons\label{tab:masscomparisons}}
\tablehead{
& \ \ \ \ \ \ \ \ \underbar{All Surveys} &  & \ \ \ \ \ \ \ \ \underbar{Dwarfs Only} &  & \ \ \ \ \ \ \ \ \underbar{SHIELD Only} &  \\
\colhead{Mass Quantity} & \colhead{Slope} & \colhead{Scatter} & \colhead{Slope} & \colhead{Scatter} & \colhead{Slope} & \colhead{Scatter} \\
(1) & (2) & (3) & (4) & (5) & (6) & (7)
}
\startdata 
\ion{H}{1} Mass & 0.816 $\pm$ 0.028 & 0.388 & $0.846 \pm 0.048$ & 0.359 & $0.466 \pm 0.074$ & 0.203\\
Stellar Mass & 0.984 $\pm$ 0.023 & 0.420 & $0.848 \pm 0.069$ & 0.427 & $0.911 \pm 0.148$ & 0.411\\
Baryonic Mass & 0.901 $\pm$ 0.023 & 0.264 & $0.836 \pm 0.041$ & 0.256 & $0.631 \pm 0.072$ & 0.203\\
\\
log(L$_B$) & 0.953 $\pm$ 0.048 & 0.349 & 0.945 $\pm$ 0.050 & 0.374 & 0.903 $\pm$ 0.138 & 0.386 \\
\enddata
\tablecomments {``All Surveys" refers to: SHIELD, FIGGS, LITTLE THINGS, VLA ANGST and 11HUGS. ``Dwarfs Only" refers to the comparison samples of which nearly all of the galaxies are dwarfs: SHIELD, FIGGS, LITTLE THINGS, and VLA ANGST.}
\end{deluxetable*}

It is worth emphasizing that the tight linear relationships seen in Figures~\ref{fig:HImasscomparisonplot} - \ref{fig:absmagplot} imply the important fact that FUV flux is a {\it robust and reliable star formation indicator over the full mass range covered by the galaxies in our combined sample}.  The same point can clearly not be said for the H$\alpha$ SFRs.  While the latter method provides accurate estimates of SFRs for more massive galaxies, it become increasingly unreliable at lower masses and lower SFRs.  This point will be explored in more detail in Shepley et al. 2026 (in preparation).

Interestingly, the overall scatter about the fits for ``All Surveys" data (column 3) is at a minimum for the baryonic mass -- FUV SFR relation.  This is consistent with the visual impression one gets by looking at Figures~\ref{fig:HImasscomparisonplot} - \ref{fig:absmagplot}, where the galaxies plotted in Figure~\ref{fig:BaryonicMassPlot} show a much tighter relation than seen in the other three plots.  At first glance, this result may seem a bit odd, since the baryonic mass is simply a linear combination of the stellar and \ion{H}{1} masses, both of which exhibit substantially larger scatter in their respective relations with FUV SFR.  What we find is that the combination of the stellar and gas masses results in substantially lower scatter, implying that the baryonic mass is the best predictor of the SFR in gas-rich galaxies (of the options we have considered).  Perhaps this result is not surprising, since previous studies have found that using the baryonic mass to construct the Tully-Fisher relation for galaxies \citep[a.k.a., the baryonic Tully-Fisher relation (BTFR);][]{begum2008a,TF2011, mcgaugh2012, lelli2016b, mcquinn2022} also results in a relation with lower scatter.

When looking at the ``Dwarfs Only" comparisons, we see that the slopes for the three mass values all cluster around 0.85, while the slope for the luminosity relation fit stays the same as was found when we used all of the comparison data.   In all cases, the uncertainties in the slopes increase, presumably because the fit is taking place over a more restrictive range of SFR values. The values of the scatter about the various fits (column 5) are essentially the same as the corresponding values for the ``All Surveys" fits.  Once again, the baryonic mass results in the smallest scatter.

For the ``SHIELD Only" fits, we see that the slopes show a much wider range of values and exhibit larger uncertainties when compared to the fits from the larger samples. When looking at the ``SHIELD Only" fits, we also see that the slopes for both the
\ion{H}{1} mass fit and the baryonic mass fit get dramatically flatter compared to the fits to the larger samples, while the stellar mass best fit slope remains nearly the same.  Again, we attribute the variations in the slopes as well as the larger uncertainties to the  small dynamic range covered by the SHIELD sample alone.  The scatter for all 4 mass-quantity fits is smaller when fitting to only SHIELD galaxies.  This is most likely due to the higher degree of homogeneity in the SHIELD sample compared to that of the various other comparison samples. In particular, we note that the scatter in the ``SHIELD Only" \ion{H}{1} mass fit is substantially reduced when compared to the ``Dwarfs Only" and ``All Surveys" \ion{H}{1} mass fits.  Once again, the scatter in the baryonic mass - FUV SFR relation is the smallest.

In summary, the baryonic mass - FUV SFR relation exhibits the smallest scatter across all of our comparisons, implying that M$_{baryonic}$ is the best mass-related tracer for predicting the FUV SFR.

\subsection{Gas Quantity Comparisons}\label{sec:Gas}

In this section we derive and evaluate two properties that link to the gaseous content of the SHIELD galaxies: gas depletion timescales and gas mass fractions.

\subsubsection{Gas Depletion Timescales}\label{sec:gdts}

\begin{figure}
\centering
\includegraphics[width=3.35in]{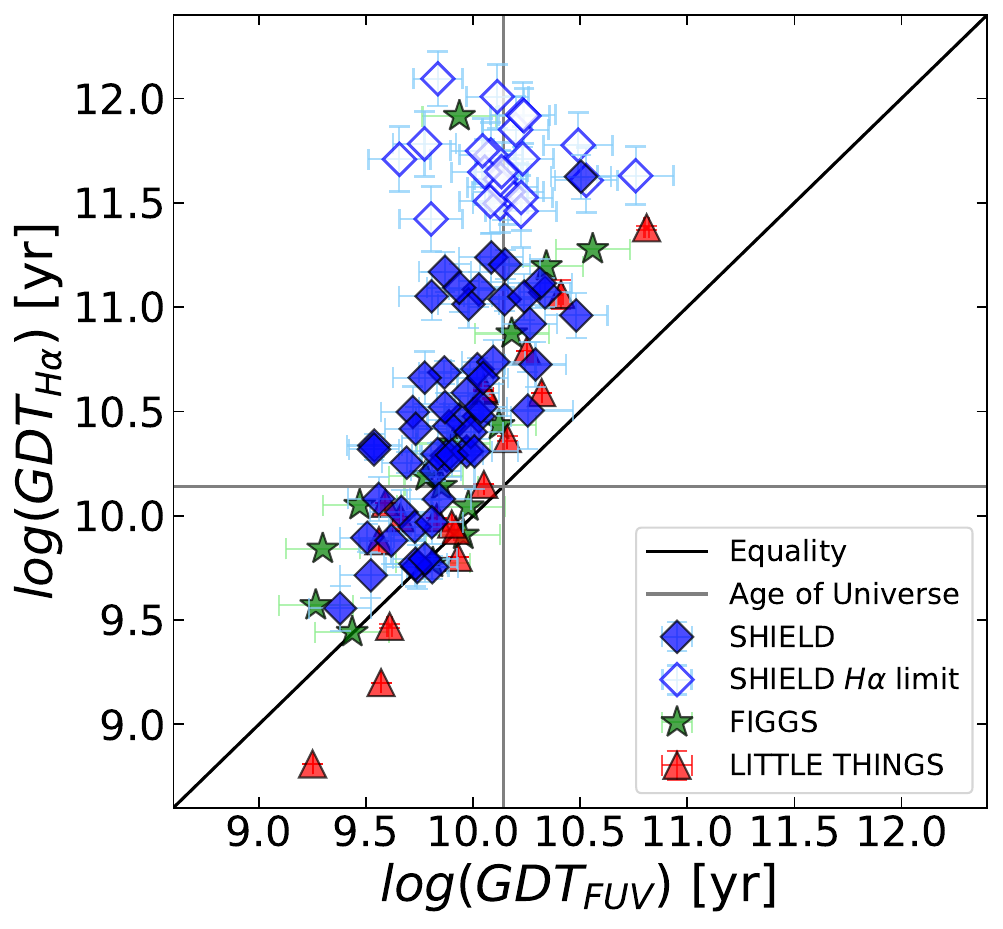}
\caption{Log-log graph of H$\alpha$-based gas depletion timescale (GDT) vs. FUV-based GDT. The solid black diagonal line represents equality between the two GDTs and the solid gray lines denote the age of the universe. }
\label{fig:GDTplot}
\end{figure}

The results of our GDT calculations are shown in Figure~\ref{fig:GDTplot}, where we plot the logarithm of the H$\alpha$-based GDT vs.\ the logarithm of the FUV-based GDT for the SHIELD galaxies as well as for the dwarf-dominated LITTLE THINGS (red triangles)  and FIGGS (green stars) comparison samples.  We point out that the comparison galaxies located in the lower left portion of the figure that are well separated from the SHIELD galaxies are actually higher mass systems that lie above the locations occupied by the SHIELD galaxies on all of the various mass - SFR plots shown above (Figures~\ref{fig:HImasscomparisonplot} - \ref{fig:absmagplot}).  With the exception of these higher-mass outliers, the comparison sample galaxies tend to fall in the same region of parameter space as the SHIELD galaxies.

The gas depletion timescale (GDT) of a galaxy is traditionally computed by dividing the total amount of available gas by the rate at which that gas is being converted into stars.  Given our previously stated caveats about the lack of molecular gas measurements in dwarf galaxies, we will compute the GDT for each SHIELD galaxy using the following equation:
\begin{equation}
  GDT = M_{gas} / SFR,
\end{equation}
where M$_{gas}$ is given by equation 6.  We compute GDTs using both the FUV and H$\alpha$ SFRs (GDT$_{FUV}$ and GDT$_{H\alpha}$, respectively).

The distribution of SHIELD galaxies in Figure~\ref{fig:GDTplot} is predictable based on what we saw in Figure~\ref{fig:SFRcomparisonplot1}. Most SHIELD galaxies lie to the left of the equality line in the GDT$_{H\alpha}$ vs. GDT$_{FUV}$ plot, since their H$\alpha$ SFRs are typically less than their FUV SFRs.  That is, GDT$_{H\alpha}$ $>>$ GDT$_{FUV}$ in most cases.  The majority of the SHIELD galaxies have H$\alpha$-based GDTs that are in excess of the age of the universe (horizontal solid line) even if the galaxies with upper limits are excluded, implying that they have used up less than half of their initial allotment of gas.   
%The average GDT$_{H\alpha}$ for the SHIELD galaxies is 15.8 Gyr if only the 55 H$\alpha$-detected galaxies are considered, and rises to 25.4 Gyr when the galaxies with upper limits on their H$\alpha$ SFRs are included in the calculation.  
In the most extreme case, one SHIELD galaxy with a measured H$\alpha$ SFR has enough gas to maintain its current rate of star formation for over 300 Gyr.  These values should all be interpreted with caution, since it seems clear that the H$\alpha$ SFRs badly underestimate reality.

%The distribution of SHIELD galaxies in Figure~\ref{fig:GDTplot} is predictable based on what we saw in Figure~\ref{fig:SFRcomparisonplot1}. Most SHIELD galaxies lie to the left of the equality line in the GDT$_{H\alpha}$ vs. GDT$_{FUV}$ plot, since their H$\alpha$ SFRs are typically less than their FUV SFRs.  That is, GDT$_{H\alpha}$ $>>$ GDT$_{FUV}$ in most cases.  The majority of the SHIELD galaxies have H$\alpha$-based GDTs that are in excess of the age of the universe (horizontal solid line) even if the galaxies with upper limits are excluded, implying that they have used up less than half of their initial allotment of gas.   The average GDT$_{H\alpha}$ for the SHIELD galaxies is 15.8 Gyr if only the 55 H$\alpha$-detected galaxies are considered, and rises to 25.4 Gyr when the galaxies with upper limits on their H$\alpha$ SFRs are included in the calculation.  In the most extreme cases, a few SHIELD galaxies with measured H$\alpha$ SFRs have enough gas to maintain their current rate of star formation for over 300 Gyr.  These values should all be interpreted with caution, since it seems clear that the H$\alpha$ SFRs badly underestimate reality.

Under the assumption that the FUV SFRs are more representative of their true values averaged over a few hundred million years, we see that the typical SHIELD galaxies have GDT$_{FUV}$ somewhat less than the age of the universe.  That is, the majority lie to the left of the vertical line in Figure~\ref{fig:GDTplot}.  The average GDT$_{FUV}$ for the 75 SHIELD galaxies with FUV fluxes is 9.3 Gyr.  This means that the typical SHIELD galaxy has used up somewhat more than half of their initial gas at this point, but will continue to be able to produce additional stars well into the future. 

It is worth noting that there is an inherent bias in the SHIELD sample towards galaxies with long gas depletion timescales. This is due to the \ion{H}{1} selection method of the ALFALFA survey. ALFALFA preferentially finds galaxies with detectable amounts of \ion{H}{1} gas while being fairly blind to their optical properties.  That is, ALFALFA does not preferentially select higher surface brightness galaxies that would exhibit higher SFRs and hence smaller GDTs. Hence, it is no surprise that the typical SHIELD galaxy will have a supply of gas that will allow it to continue making stars for many Gyrs. This same bias exists in many of the comparison galaxies as well, but as a group the SHIELD galaxies have low SFRs to go along with their low \ion{H}{1} masses.  Only a relatively few of the SHIELD galaxies would fall in the category of ``highly active" star-forming dwarfs (see Section~\ref{sec:StellarsSFR} below).

\subsubsection{Baryonic Gas Mass Fraction vs.\ FUV SFR}\label{sec:BaryonicVsSFR}

\begin{figure}
\centering
\includegraphics[width=3.35in]{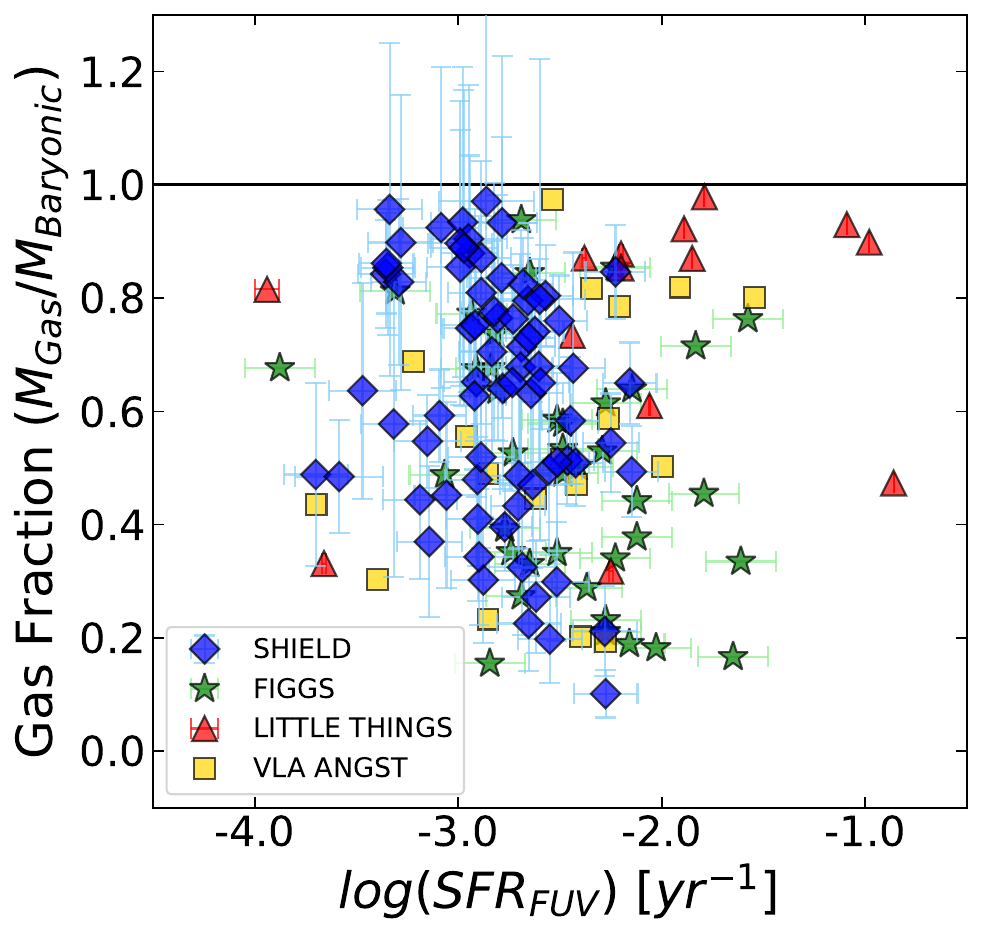}
\caption{Graph of baryonic gas mass fraction (M$_{gas}$ / M$_{baryonic}$) vs. the log of FUV SFR.  The horizontal solid line indicates M$_{gas}$ = M$_{baryonic}$, which would be the case for a starless galaxy.}
\label{fig:BaryonicFractionVssfr}
\end{figure}

We next consider the baryonic gas mass fraction (M$_{gas}$ / M$_{baryonic}$) for the SHIELD galaxies and compare it to the FUV SFRs.  The results are shown in Figure~\ref{fig:BaryonicFractionVssfr}.

\begin{figure}
\centering
\includegraphics[width=3.35in]{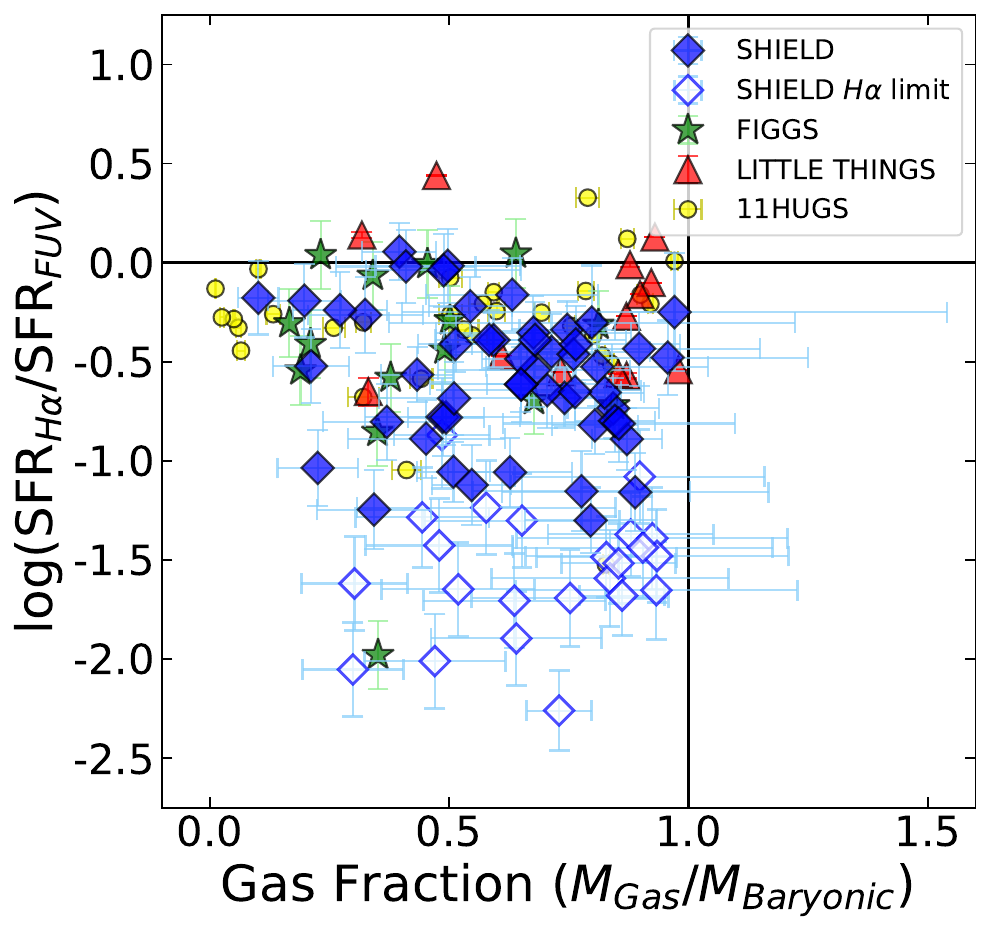}
\caption{Graph of the log of (H$\alpha$ SFR/FUV SFR) vs.\ baryonic gas mass fraction (M$_{gas}$ / M$_{baryonic}$). The vertical solid line indicates M$_{gas}$ = M$_{baryonic}$ (starless galaxies), and the horizontal solid line indicates SFR$_{H\alpha}$/SFR$_{FUV}$ = 1.}
\label{fig:RatioVsBaryonicFractionPlot}
\end{figure}

A basic assumption that one might make is that a galaxy's SFR depends on its gas fraction, since galaxies with higher gas content might be expected to be more effective at turning their gas into stars. However, we see in Figure~\ref{fig:BaryonicFractionVssfr}, which includes the SHIELD plus comparison sample dwarf galaxies, that this is not the case.  At any given gas mass fraction, the galaxies are spread across over two orders of magnitude in the FUV SFR with no apparent trends with gas fraction.  The SHIELD galaxies cover a somewhat smaller range in FUV SFR, but again show no obvious trend with gas fraction.  These results add further support to the conclusions drawn from Section~\ref{sec:HI} that the process which governs star formation in these gas-rich dwarfs is more complex than just the amount of available gas.  

One of the interesting aspects of the SHIELD sample illustrated in Figure~\ref{fig:BaryonicFractionVssfr} is the fact that many of these galaxies have very large gas mass fractions.  While a typical spiral galaxy might have M$_{gas}$ / M$_{baryonic}$ $\sim$ 0.1 or less, there are many dwarfs in both the SHIELD and comparison samples that have M$_{gas}$ / M$_{baryonic}$ $>$ 0.8 (i.e., M$_{gas}$ $>$ 4$\cdot$M$_{stellar}$).  Fully 68.3\% of the SHIELD galaxies (56 out of 82) have M$_{gas}$ / M$_{baryonic}$ $>$ 0.5 (i.e., M$_{gas}$ $>$ M$_{stellar}$).   Hence, the SHIELD galaxies are not just a gas-rich sample of dwarfs, they are a gas-{\it dominated} sample.

The conclusion that the gas mass fraction does not correlate with the star-forming characteristics of the SHIELD galaxies (or with dwarf galaxies in general) is further supported by Figure~\ref{fig:RatioVsBaryonicFractionPlot}, where we compare the SFR ratio (SFR$_{H\alpha}$ / SFR$_{FUV}$) to the baryonic gas mass fraction.  This figure shows that the ratio of SFR$_{H\alpha}$ / SFR$_{FUV}$ does not depend on the gas fraction.  As was the case with Figure~\ref{fig:BaryonicFractionVssfr}, we see that at any given value of M$_{gas}$ / M$_{baryonic}$ the SFR ratio covers the full range of observed values between $-$2.0 and 0.0.  The only exceptions are a handful of high mass 11HUGS galaxies on the left side of Figure~\ref{fig:RatioVsBaryonicFractionPlot} with low gas fractions and SFR ratios between $-$0.5 and 0.0.  Whatever mechanism is driving the differences in the H$\alpha$ and FUV SFRs seen in Figures~\ref{fig:SFRcomparisonplot1} and~\ref{fig:SFRcomparisonplot2} does not appear to be linked to the gas fraction of the galaxies.

\subsection{Specific Star-Formation Rate Comparisons}\label{sec:sSFR}

The specific star-formation rate (sSFR), defined by $(SFR/M_{Stellar})$, is a useful metric for determining how efficiently a galaxy creates stars.  Because this quantity is mass normalized, it also allows for an easier comparison between galaxy samples that cover a range of masses. %Thus, we have included 11HUGS in Figure~\ref{fig:StellarMassVsssfr} and Figure~\ref{fig:BaryonicFractionVsssfr}. 

\subsubsection{Stellar Mass vs.\ FUV sSFR}\label{sec:StellarsSFR}
\begin{figure}
\centering
\includegraphics[width=3.35in]{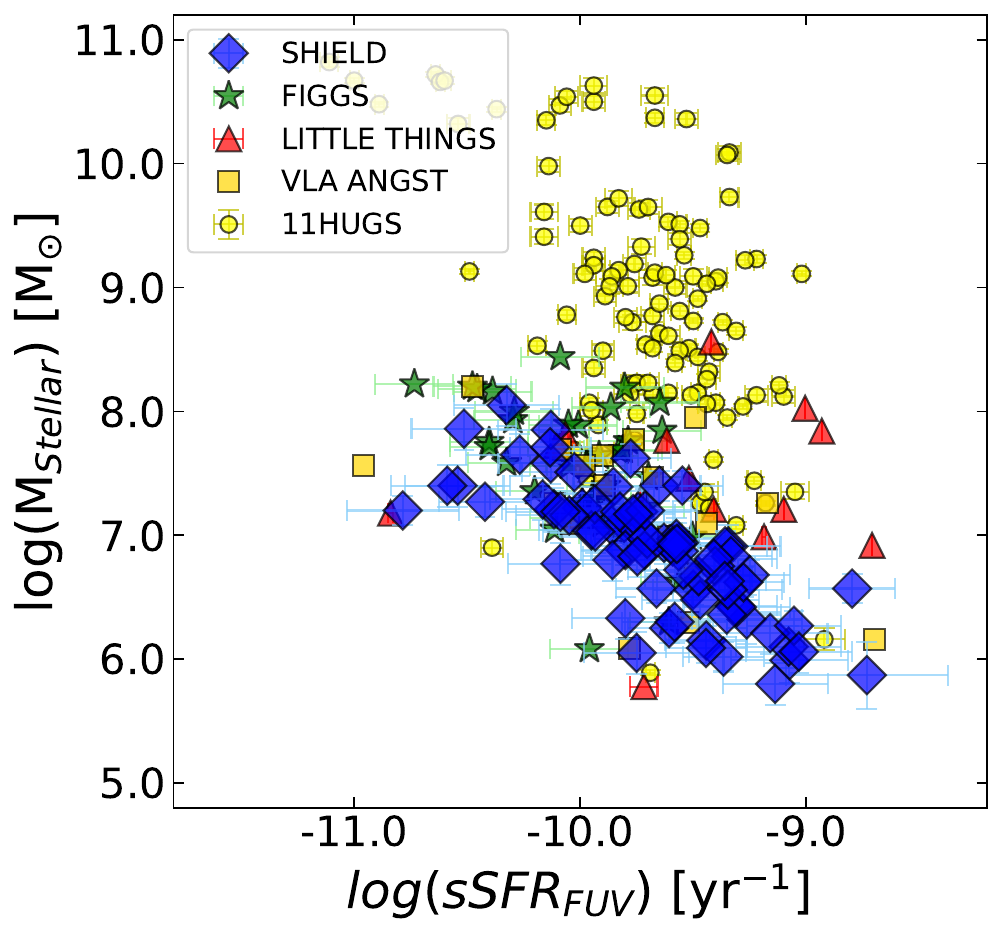}
\caption{Log-log graph of Stellar Mass vs.\ FUV sSFR.  In addition to the SHIELD galaxies we include the comparison samples from FIGGS, LITTLE THINGS, VLA ANGST, and 11HUGs.  The SHIELD galaxies anchor the low-mass end of plot.}
\label{fig:StellarMassVsssfr}
\end{figure}

We compare stellar mass to the FUV sSFR in Figure~\ref{fig:StellarMassVsssfr}.   This plot has the advantage of allowing us to look for any trends in the sSFR between the dwarfs and the more massive galaxies, represented here primarily by the 11HUGS sample.  The bulk of the SHIELD sample have FUV sSFR values between $-$9 and $-$11, similar to the more massive 11HUGS galaxies.  The dwarf comparison sample galaxies from FIGGS, LITTLE THINGS and VLA ANGST overlap with the SHIELD galaxies at the lower masses, with the LITTLE THINGS galaxies tending to exhibit higher sSFR values (to the right in Figure~\ref{fig:StellarMassVsssfr}), while the FIGGS galaxies tend to lie to lower sSFR values. This is presumably a reflection of the different selection methods used to create these two samples of dwarf galaxies.

While there are no strong trends exhibited by any of  the comparison samples in Figure~\ref{fig:StellarMassVsssfr}, for the SHIELD galaxies we do see a clear trend of increasing FUV sSFR as stellar mass decreases.  We attribute this apparent trend to the fact that the SHIELD galaxies occupy a rather limited range in both M$_{Stellar}$ and FUV SFR (e.g., Figure~\ref{fig:stellarmassvssfr}), and not to any special characteristics of the SHIELD sample.

Despite their high gas mass fractions (e.g., Figure~\ref{fig:BaryonicFractionVssfr}), Figure~\ref{fig:StellarMassVsssfr} reveals that many of the SHIELD galaxies are relatively inefficient at forming stars.  This low efficiency is consistent with their high GDTs shown in Figure~\ref{fig:GDTplot}.   We speculate that these low star-formation efficiencies are linked to the fact that most SHIELD galaxies exhibit peak \ion{H}{1} column densities below 10$^{21}$ cm$^{-2}$ \citep{cannon2011, teich2016}.  This surface density of \ion{H}{1} gas is usually considered to be the lower limit necessary for the formation of giant \ion{H}{2} region complexes \citep[e.g.,][]{skillman1987, kennicutt1998sl}.  A comparison between existing \ion{H}{1} maps and H$\alpha$ images \citep{teich2016} does reveal some current star formation at lower column densities (a few times 10$^{20}$ cm$^{-2}$) in many of the SHIELD galaxies, but it is clear that much of the gas is not participating in star-formation activity.  We will return to this important point in future papers in this series.

There are only a small number of galaxies with mass-normalized SFRs above $-$9.0, which would categorize them as exhibiting strong star formation activity (e.g., characteristic of blue compact dwarf galaxies).  One of these is AGC 198691 (a.k.a. the Leoncino dwarf), which is the rightmost of the SHIELD galaxies in Figure~\ref{fig:StellarMassVsssfr} (log(sSFR) = $-$8.73 yr$^{-1}$, log(M$_{stellar}$) = 5.87 M$_\odot$).  This galaxy is notable for its low stellar mass and extremely low metal abundance \citep[log(O/H)+12 = 7.06 $\pm$ 0.03,][]{hirschauer2016, aver2022}.  Galaxies with even higher sSFR values (e.g., $-$8.5 to $-$7.0) are almost exclusively extreme star-forming systems such as Green Pea \citep{cardamone2009, brunker2020} or Blueberry \citep{yang2017} galaxies \citep[see Figure 4 of][] {hirschauer2022}.  There are no examples of such extreme star-forming galaxies in SHIELD or any of our comparison samples, reflecting the rarity of galaxies characterized by strong, global bursts of star formation \citep[e.g.,][]{jclee2009, brunker2020}.

\subsubsection{Baryonic Gas Mass Fraction vs.\ sSFR}\label{sec:BaryonicFracVssSFR}

\begin{figure}
\centering
\includegraphics[width=3.35in]{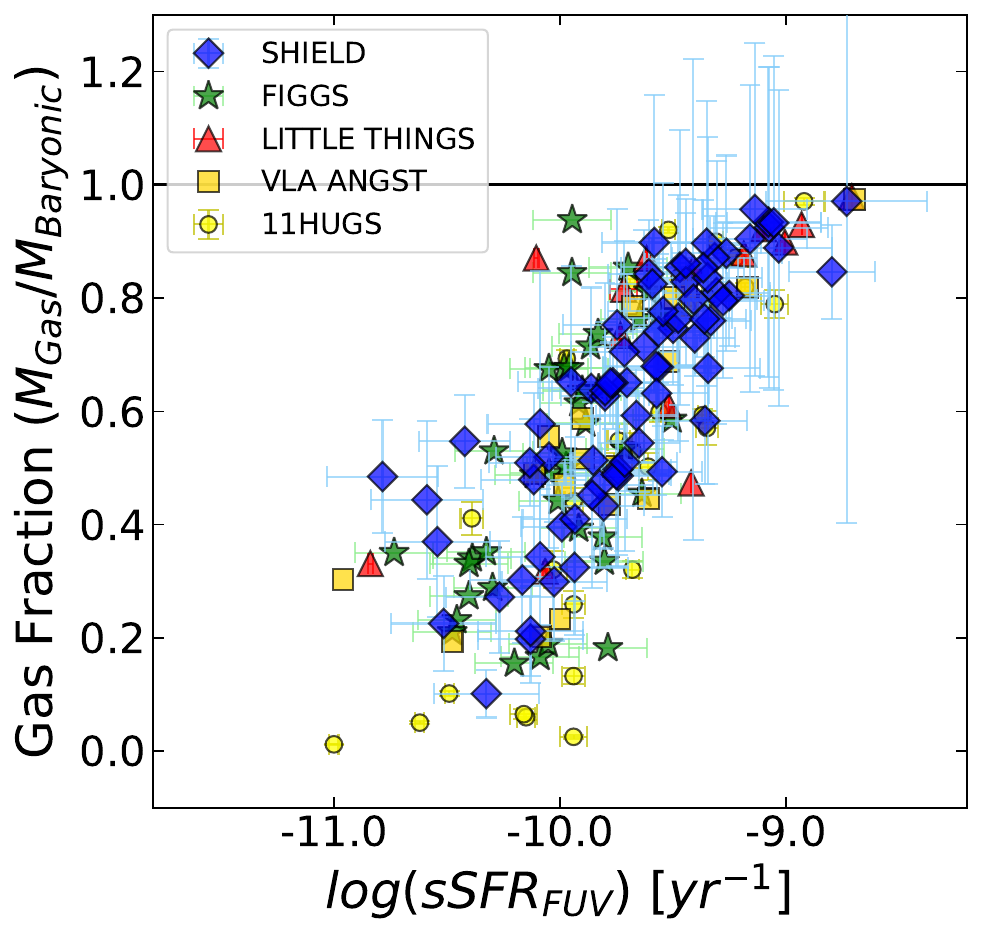}
\caption{Graph of baryonic gas mass fraction $(M_{gas} / M_{baryonic})$ vs.\ the log of FUV sSFR. The solid line indicates equality $(M_{gas} = M_{baryonic})$.}
\label{fig:BaryonicFractionVsssfr}
\end{figure}

\vskip 0.4inWe compare the baryonic gas mass fraction $(M_{gas} / M_{baryonic})$ to the FUV sSFR to gain a better understanding of the interplay between gas availability and star formation efficiency in dwarf galaxies. The results are shown in Figure~\ref{fig:BaryonicFractionVsssfr}, where some interesting trends are evident.

First, we see that the SHIELD galaxies overlap the comparison samples thoroughly.  This is distinct from the distribution of the galaxies in Figure~\ref{fig:StellarMassVsssfr} (which share the same X-axis but where the Y-axis is now represented by a mass-normalized quantity) and Figure~\ref{fig:BaryonicFractionVssfr} (which share the same Y-axis but where the X-axis is now represented by a mass-normalized quantity).   In these previous plots the SHIELD galaxies were often offset from members of the comparison samples that possess higher masses or higher SFRs, while in Figure~\ref{fig:BaryonicFractionVsssfr} these offsets disappear and all of the galaxies fall within the same band.

More importantly, the galaxies plotted in Figure~\ref{fig:BaryonicFractionVsssfr} show a significant trend of increasing FUV sSFR with increasing baryonic gas mass fraction.  Galaxies with the highest sSFR values are limited to galaxies with high gas mass fractions.  Conversely, the galaxies with the lowest gas mass fractions are found among those galaxies with lower mass-normalized SFRs.  To be sure, there is significant scatter in the relation evident in the plot ($\sim$0.4 dex spread in log(sSFR$_{FUV}$) for galaxies with gas mass fractions below 0.6 and $\sim$0.2 with larger gas mass fractions), yet the spread in the sSFR values at a fixed gas fraction value is significantly lower than the corresponding scatter seen in Figure~\ref{fig:BaryonicFractionVssfr}.

%************************************************************
%\section{Discussion}\label{sec:Discussion}
%\subsection{SFR vs Mass Correlations}\label{sec:MassDiscussion}

\vskip 0.4in
\section{Summary and Conclusions}\label{sec:summary}
SHIELD is a  multi-wavelength observational program targeting gas-rich, star-forming dwarf galaxies at the faint end of the \ion{H}{1} mass function. The survey is volumetrically complete within the given selection constraints, yielding a highly representative sample of galaxies within this mass range. The sample includes systems that are among the least massive star-forming galaxies known in the local universe, making them attractive targets for studying galaxy formation and evolution at the lowest-masses.

This paper is the first published analysis that includes the complete SHIELD sample of 82 galaxies. We have presented FUV fluxes and SFRs for 75 of 82 SHIELD galaxies.  Of the seven galaxies missing FUV SFRs, six galaxies had no GALEX images and one galaxy was too faint in its shallow GALEX image to make a reliable measurement. 

The main results of this paper can be summarized as follows:
\begin{itemize}

\item The SHIELD galaxies add substantially to our knowledge of galaxy properties at lower masses.  For example, they act to fill in and better define the relation between the H$\alpha$ and FUV SFRs for dwarf galaxies.
%region around the departure from the H$\alpha$ SFR to FUV SFR equality in Figures~\ref{fig:SFRcomparisonplot1} and~\ref{fig:SFRcomparisonplot2}.  
The results of comparing the H$\alpha$ and FUV SFRs support the idea of the highly stochastic nature of star formation in extremely low-mass galaxies.

\item We explore the connections between FUV SFR and \ion{H}{1} mass, stellar mass, baryonic mass, and B-band absolute magnitude.  Combining the SHIELD galaxies with an extensive comparison sample of nearby, gas-rich galaxies, we find strong correlations between FUV SFR and all four of these mass-like quantities.   We find that the baryonic mass of a galaxy is the best mass-related tracer of its FUV SFR. 

\item Furthermore, our analysis of FUV SFRs and galaxy masses reconfirms the recognition that FUV data provide robust estimates of star formation over the full range of galaxy masses covered by this study, including the very lowest mass gas-rich dwarfs.

\item The gas depletion timescales (GDTs) of SHIELD galaxies are long, and indicate that the typical SHIELD galaxy has used up only half of its available gas.  The majority of SHIELD galaxies (56 of 82) have baryonic gas mass fractions over 0.5 (i.e., have M$_{gas}$ $>$ M$_{stellar}$), meaning that they are gas dominated systems.

\item Based on their observed sSFR values and consistent with their long GDTs, the typical SHIELD galaxy does not make stars very efficiently.  It may be that these low star-formation efficiencies are due to the relatively low peak \ion{H}{1} column densities observed in most SHIELD galaxies.

\end{itemize}

A companion paper will publish a complete description of our H$\alpha$ measurements, calculation methodology, and results for the full SHIELD sample (Shepley et al. 2026, in preparation).

%************************************************************
%\section{acknowledgements}

\begin{acknowledgements}

We thank the anonymous referee for many helpful suggestions and corrections.  DG would like to thank John and A-Lan Reynolds for their support of the Reynolds Post-Baccalaureate Fellowship, administered by the Indiana University Astronomy Department, which supported him during the completion of this project.  JJS would like to express his thanks and gratitude to the College of Arts and Sciences at Indiana University for their ongoing support of research carried out in the Astronomy Department.  This project utilized archival images obtained by the GALEX satellite and made available to the community through the Mikulski Archive for Space Telescopes (MAST). We gratefully acknowledge the many scientists who worked to create and operate GALEX as well as the members of the MAST team who continue to make these valuable data available.

\end{acknowledgements}

\vspace{5mm}
\facilities{GALEX, MAST}

%%pdflatex sample631.tex
%%bibtext sample631
%%pdflatex sample631.tex
%%pdflatex sample631.tex

\bibliographystyle{aasjournal}
\bibliography{SHIELD_UV_SFR}{}

@ARTICLE{cannon2011,
       author = {{Cannon}, John M. and {Giovanelli}, Riccardo and {Haynes}, Martha P. and {Janowiecki}, Steven and {Parker}, Angela and {Salzer}, John J. and {Adams}, Elizabeth A.~K. and {Engstrom}, Eric and {Huang}, Shan and {McQuinn}, Kristen B.~W. and {Ott}, J{\"u}rgen and {Saintonge}, Am{\'e}lie and {Skillman}, Evan D. and {Allan}, John and {Erny}, Grace and {Fliss}, Palmer and {Smith}, AnnaLeigh},
        title = "{The Survey of H I in Extremely Low-mass Dwarfs (SHIELD)}",
      journal = {\apjl},
         year = 2011,
        month = sep,
       volume = {739},
       number = {1},
          eid = {L22},
        pages = {L22},
          doi = {10.1088/2041-8205/739/1/L22},
archivePrefix = {arXiv},
       eprint = {1105.4505},
 primaryClass = {astro-ph.CO},
       adsurl = {https://ui.adsabs.harvard.edu/abs/2011ApJ...739L..22C}
}

@ARTICLE{hirschauer2016,
       author = {{Hirschauer}, Alec S. and {Salzer}, John J. and {Skillman}, Evan D. and {Berg}, Danielle and {McQuinn}, Kristen B.~W. and {Cannon}, John M. and {Gordon}, Alex J.~R. and {Haynes}, Martha P. and {Giovanelli}, Riccardo and {Adams}, Elizabeth A.~K. and {Janowiecki}, Steven and {Rhode}, Katherine L. and {Pogge}, Richard W. and {Croxall}, Kevin V. and {Aver}, Erik},
        title = "{ALFALFA Discovery of the Most Metal-poor Gas-rich Galaxy Known: AGC 198691}",
      journal = {\apj},
         year = 2016,
        month = may,
       volume = {822},
       number = {2},
          eid = {108},
        pages = {108},
          doi = {10.3847/0004-637X/822/2/108},
archivePrefix = {arXiv},
       eprint = {1603.03798},
 primaryClass = {astro-ph.GA},
       adsurl = {https://ui.adsabs.harvard.edu/abs/2016ApJ...822..108H}
}

@ARTICLE{aver2022,
       author = {{Aver}, Erik and {Berg}, Danielle A. and {Hirschauer}, Alec S. and {Olive}, Keith A. and {Pogge}, Richard W. and {Rogers}, Noah S.~J. and {Salzer}, John J. and {Skillman}, Evan D.},
        title = "{A comprehensive chemical abundance analysis of the extremely metal poor Leoncino Dwarf galaxy (AGC 198691)}",
      journal = {\mnras},
         year = 2022,
        month = feb,
       volume = {510},
       number = {1},
        pages = {373-382},
          doi = {10.1093/mnras/stab3226},
archivePrefix = {arXiv},
       eprint = {2109.00178},
 primaryClass = {astro-ph.GA},
       adsurl = {https://ui.adsabs.harvard.edu/abs/2022MNRAS.510..373A}
}

@ARTICLE{haurberg2015,
       author = {{Haurberg}, Nathalie C. and {Salzer}, John J. and {Cannon}, John M. and {Marshall}, Melissa V.},
        title = "{Oxygen Abundance Measurements of SHIELD Galaxies}",
      journal = {\apj},
         year = 2015,
        month = feb,
       volume = {800},
       number = {2},
          eid = {121},
        pages = {121},
          doi = {10.1088/0004-637X/800/2/121},
archivePrefix = {arXiv},
       eprint = {1501.05159},
 primaryClass = {astro-ph.GA},
       adsurl = {https://ui.adsabs.harvard.edu/abs/2015ApJ...800..121H}
}

@ARTICLE{haynes2011,
       author = {{Haynes}, Martha P. and {Giovanelli}, Riccardo and {Martin}, Ann M. and {Hess}, Kelley M. and {Saintonge}, Am{\'e}lie and {Adams}, Elizabeth A.~K. and {Hallenbeck}, Gregory and {Hoffman}, G. Lyle and {Huang}, Shan and {Kent}, Brian R. and {Koopmann}, Rebecca A. and {Papastergis}, Emmanouil and {Stierwalt}, Sabrina and {Balonek}, Thomas J. and {Craig}, David W. and {Higdon}, Sarah J.~U. and {Kornreich}, David A. and {Miller}, Jeffrey R. and {O'Donoghue}, Aileen A. and {Olowin}, Ronald P. and {Rosenberg}, Jessica L. and {Spekkens}, Kristine and {Troischt}, Parker and {Wilcots}, Eric M.},
        title = "{The Arecibo Legacy Fast ALFA Survey: The {\ensuremath{\alpha}}.40 H I Source Catalog, Its Characteristics and Their Impact on the Derivation of the H I Mass Function}",
      journal = {\aj},
         year = 2011,
        month = nov,
       volume = {142},
       number = {5},
          eid = {170},
        pages = {170},
          doi = {10.1088/0004-6256/142/5/170},
archivePrefix = {arXiv},
       eprint = {1109.0027},
 primaryClass = {astro-ph.CO},
       adsurl = {https://ui.adsabs.harvard.edu/abs/2011AJ....142..170H}
}

@ARTICLE{haynes2018,
       author = {{Haynes}, Martha P. and {Giovanelli}, Riccardo and {Kent}, Brian R. and {Adams}, Elizabeth A.~K. and {Balonek}, Thomas J. and {Craig}, David W. and {Fertig}, Derek and {Finn}, Rose and {Giovanardi}, Carlo and {Hallenbeck}, Gregory and {Hess}, Kelley M. and {Hoffman}, G. Lyle and {Huang}, Shan and {Jones}, Michael G. and {Koopmann}, Rebecca A. and {Kornreich}, David A. and {Leisman}, Lukas and {Miller}, Jeffrey and {Moorman}, Crystal and {O'Connor}, Jessica and {O'Donoghue}, Aileen and {Papastergis}, Emmanouil and {Troischt}, Parker and {Stark}, David and {Xiao}, Li},
        title = "{The Arecibo Legacy Fast ALFA Survey: The ALFALFA Extragalactic H I Source Catalog}",
      journal = {\apj},
         year = 2018,
        month = jul,
       volume = {861},
       number = {1},
          eid = {49},
        pages = {49},
          doi = {10.3847/1538-4357/aac956},
archivePrefix = {arXiv},
       eprint = {1805.11499},
 primaryClass = {astro-ph.GA},
       adsurl = {https://ui.adsabs.harvard.edu/abs/2018ApJ...861...49H}
}

@ARTICLE{giovanelli2005,
       author = {{Giovanelli}, Riccardo and {Haynes}, Martha P. and {Kent}, Brian R. and {Perillat}, Philip and {Saintonge}, Amelie and {Brosch}, Noah and {Catinella}, Barbara and {Hoffman}, G. Lyle and {Stierwalt}, Sabrina and {Spekkens}, Kristine and {Lerner}, Mikael S. and {Masters}, Karen L. and {Momjian}, Emmanuel and {Rosenberg}, Jessica L. and {Springob}, Christopher M. and {Boselli}, Alessandro and {Charmandaris}, Vassilis and {Darling}, Jeremy K. and {Davies}, Jonathan and {Garcia Lambas}, Diego and {Gavazzi}, Giuseppe and {Giovanardi}, Carlo and {Hardy}, Eduardo and {Hunt}, Leslie K. and {Iovino}, Angela and {Karachentsev}, Igor D. and {Karachentseva}, Valentina E. and {Koopmann}, Rebecca A. and {Marinoni}, Christian and {Minchin}, Robert and {Muller}, Erik and {Putman}, Mary and {Pantoja}, Carmen and {Salzer}, John J. and {Scodeggio}, Marco and {Skillman}, Evan and {Solanes}, Jose M. and {Valotto}, Carlos and {van Driel}, Wim and {van Zee}, Liese},
        title = "{The Arecibo Legacy Fast ALFA Survey. I. Science Goals, Survey Design, and Strategy}",
      journal = {\aj},
         year = 2005,
        month = dec,
       volume = {130},
       number = {6},
        pages = {2598-2612},
          doi = {10.1086/497431},
archivePrefix = {arXiv},
       eprint = {astro-ph/0508301},
 primaryClass = {astro-ph},
       adsurl = {https://ui.adsabs.harvard.edu/abs/2005AJ....130.2598G}
}

@ARTICLE{teich2016,
       author = {{Teich}, Yaron G. and {McNichols}, Andrew T. and {Nims}, Elise and {Cannon}, John M. and {Adams}, Elizabeth A.~K. and {Giovanelli}, Riccardo and {Haynes}, Martha P. and {McQuinn}, Kristen B.~W. and {Salzer}, John J. and {Skillman}, Evan D. and {Bernstein-Cooper}, Elijah Z. and {Dolphin}, Andrew and {Elson}, E.~C. and {Haurberg}, Nathalie and {J{\'o}zsa}, Gyula I.~G. and {Ott}, J{\"u}rgen and {Saintonge}, Amelie and {Warren}, Steven R. and {Cave}, Ian and {Hagen}, Cedric and {Huang}, Shan and {Janowiecki}, Steven and {Marshall}, Melissa V. and {Thomann}, Clara M. and {Van Sistine}, Angela},
        title = "{SHIELD: Comparing Gas and Star Formation in Low-mass Galaxies}",
      journal = {\apj},
         year = 2016,
        month = nov,
       volume = {832},
       number = {1},
          eid = {85},
        pages = {85},
          doi = {10.3847/0004-637X/832/1/85},
archivePrefix = {arXiv},
       eprint = {1609.05375},
 primaryClass = {astro-ph.GA},
       adsurl = {https://ui.adsabs.harvard.edu/abs/2016ApJ...832...85T}
}

@ARTICLE{mcnichols2016,
       author = {{McNichols}, Andrew T. and {Teich}, Yaron G. and {Nims}, Elise and {Cannon}, John M. and {Adams}, Elizabeth A.~K. and {Bernstein-Cooper}, Elijah Z. and {Giovanelli}, Riccardo and {Haynes}, Martha P. and {J{\'o}zsa}, Gyula I.~G. and {McQuinn}, Kristen B.~W. and {Salzer}, John J. and {Skillman}, Evan D. and {Warren}, Steven R. and {Dolphin}, Andrew and {Elson}, E.~C. and {Haurberg}, Nathalie and {Ott}, J{\"u}rgen and {Saintonge}, Amelie and {Cave}, Ian and {Hagen}, Cedric and {Huang}, Shan and {Janowiecki}, Steven and {Marshall}, Melissa V. and {Thomann}, Clara M. and {Van Sistine}, Angela},
        title = "{SHIELD: Neutral Gas Kinematics and Dynamics}",
      journal = {\apj},
         year = 2016,
        month = nov,
       volume = {832},
       number = {1},
          eid = {89},
        pages = {89},
          doi = {10.3847/0004-637X/832/1/89},
archivePrefix = {arXiv},
       eprint = {1609.05376},
 primaryClass = {astro-ph.GA},
       adsurl = {https://ui.adsabs.harvard.edu/abs/2016ApJ...832...89M}
}

@ARTICLE{mcquinn2014,
       author = {{McQuinn}, Kristen B.~W. and {Cannon}, John M. and {Dolphin}, Andrew E. and {Skillman}, Evan D. and {Salzer}, John J. and {Haynes}, Martha P. and {Adams}, Elizabeth and {Cave}, Ian and {Elson}, Ed C. and {Giovanelli}, Riccardo and {Ott}, Ju{\"e}rgen and {Saintonge}, Am{\'e}lie},
        title = "{Distance Determinations to SHIELD Galaxies from Hubble Space Telescope Imaging}",
      journal = {\apj},
         year = 2014,
        month = apr,
       volume = {785},
       number = {1},
          eid = {3},
        pages = {3},
          doi = {10.1088/0004-637X/785/1/3},
archivePrefix = {arXiv},
       eprint = {1402.3723},
 primaryClass = {astro-ph.GA},
       adsurl = {https://ui.adsabs.harvard.edu/abs/2014ApJ...785....3M}
}

@ARTICLE{mcquinn2015a,
       author = {{McQuinn}, Kristen. B.~W. and {Cannon}, John M. and {Dolphin}, Andrew E. and {Skillman}, Evan D. and {Haynes}, Martha P. and {Simones}, Jacob E. and {Salzer}, John J. and {Adams}, Elizabeth A.~K. and {Elson}, Ed C. and {Giovanelli}, Riccardo and {Ott}, J{\"u}rgen},
        title = "{Characterizing the Star Formation of the Low-mass Shield Galaxies from Hubble Space Telescope Imaging}",
      journal = {\apj},
         year = 2015,
        month = mar,
       volume = {802},
       number = {1},
          eid = {66},
        pages = {66},
          doi = {10.1088/0004-637X/802/1/66},
archivePrefix = {arXiv},
       eprint = {1501.07313},
 primaryClass = {astro-ph.GA},
       adsurl = {https://ui.adsabs.harvard.edu/abs/2015ApJ...802...66M}
}

@ARTICLE{mcquinn2021,
       author = {{McQuinn}, Kristen B.~W. and {Telidevara}, Anjana K. and {Fuson}, Jackson and {Adams}, Elizabeth A.~K. and {Cannon}, John M. and {Skillman}, Evan D. and {Dolphin}, Andrew E. and {Haynes}, Martha P. and {Rhode}, Katherine L. and {Salzer}, John. J. and {Giovanelli}, Riccardo and {Gordon}, Alex J.~R.},
        title = "{Galaxy Properties at the Faint End of the H I Mass Function}",
      journal = {\apj},
         year = 2021,
        month = sep,
       volume = {918},
       number = {1},
          eid = {23},
        pages = {23},
          doi = {10.3847/1538-4357/ac03ae},
archivePrefix = {arXiv},
       eprint = {2105.05100},
 primaryClass = {astro-ph.GA},
       adsurl = {https://ui.adsabs.harvard.edu/abs/2021ApJ...918...23M}
}

@ARTICLE{hunter2012,
       author = {{Hunter}, Deidre A. and {Ficut-Vicas}, Dana and {Ashley}, Trisha and {Brinks}, Elias and {Cigan}, Phil and {Elmegreen}, Bruce G. and {Heesen}, Volker and {Herrmann}, Kimberly A. and {Johnson}, Megan and {Oh}, Se-Heon and {Rupen}, Michael P. and {Schruba}, Andreas and {Simpson}, Caroline E. and {Walter}, Fabian and {Westpfahl}, David J. and {Young}, Lisa M. and {Zhang}, Hong-Xin},
        title = "{Little Things}",
      journal = {\aj},
         year = 2012,
        month = nov,
       volume = {144},
       number = {5},
          eid = {134},
        pages = {134},
          doi = {10.1088/0004-6256/144/5/134},
archivePrefix = {arXiv},
       eprint = {1208.5834},
 primaryClass = {astro-ph.GA},
       adsurl = {https://ui.adsabs.harvard.edu/abs/2012AJ....144..134H}
}

@ARTICLE{ott2012,
       author = {{Ott}, J{\"u}rgen and {Stilp}, Adrienne M. and {Warren}, Steven R. and {Skillman}, Evan D. and {Dalcanton}, Julianne J. and {Walter}, Fabian and {de Blok}, W.~J.~G. and {Koribalski}, B{\"a}rbel and {West}, Andrew A.},
        title = "{VLA-ANGST: A High-resolution H I Survey of Nearby Dwarf Galaxies}",
      journal = {\aj},
         year = 2012,
        month = oct,
       volume = {144},
       number = {4},
          eid = {123},
        pages = {123},
          doi = {10.1088/0004-6256/144/4/123},
archivePrefix = {arXiv},
       eprint = {1208.3737},
 primaryClass = {astro-ph.CO},
       adsurl = {https://ui.adsabs.harvard.edu/abs/2012AJ....144..123O}
}

@ARTICLE{begum2008,
       author = {{Begum}, Ayesha and {Chengalur}, Jayaram N. and {Karachentsev}, I.~D. and {Sharina}, M.~E. and {Kaisin}, S.~S.},
        title = "{FIGGS: Faint Irregular Galaxies GMRT Survey - overview, observations and first results}",
      journal = {\mnras},
         year = 2008,
        month = may,
       volume = {386},
       number = {3},
        pages = {1667-1682},
          doi = {10.1111/j.1365-2966.2008.13150.x},
archivePrefix = {arXiv},
       eprint = {0802.3982},
 primaryClass = {astro-ph},
       adsurl = {https://ui.adsabs.harvard.edu/abs/2008MNRAS.386.1667B}
}

@ARTICLE{hirschauer2022,
       author = {{Hirschauer}, Alec S. and {Salzer}, John J. and {Haurberg}, Nathalie and {Gronwall}, Caryl and {Janowiecki}, Steven},
        title = "{H{\ensuremath{\alpha}} Dots: Direct-method Metal Abundances of Low-luminosity Star-forming Systems}",
      journal = {\apj},
         year = 2022,
        month = feb,
       volume = {925},
       number = {2},
          eid = {131},
        pages = {131},
          doi = {10.3847/1538-4357/ac402a},
archivePrefix = {arXiv},
       eprint = {2112.02135},
 primaryClass = {astro-ph.GA},
       adsurl = {https://ui.adsabs.harvard.edu/abs/2022ApJ...925..131H}
}

@ARTICLE{martin2010,
       author = {{Martin}, Ann M. and {Papastergis}, Emmanouil and {Giovanelli}, Riccardo and {Haynes}, Martha P. and {Springob}, Christopher M. and {Stierwalt}, Sabrina},
        title = "{The Arecibo Legacy Fast ALFA Survey. X. The H I Mass Function and {\ensuremath{\Omega}}\_H I from the 40\% ALFALFA Survey}",
      journal = {\apj},
         year = 2010,
        month = nov,
       volume = {723},
       number = {2},
        pages = {1359-1374},
          doi = {10.1088/0004-637X/723/2/1359},
archivePrefix = {arXiv},
       eprint = {1008.5107},
 primaryClass = {astro-ph.CO},
       adsurl = {https://ui.adsabs.harvard.edu/abs/2010ApJ...723.1359M}
}

@ARTICLE{brunker2020,
       author = {{Brunker}, Samantha W. and {Salzer}, John J. and {Janowiecki}, Steven and {Finn}, Rose A. and {Helou}, George},
        title = "{Properties of the KISS Green Pea Galaxies}",
      journal = {\apj},
         year = 2020,
        month = jul,
       volume = {898},
       number = {1},
          eid = {68},
        pages = {68},
          doi = {10.3847/1538-4357/ab9ec0},
archivePrefix = {arXiv},
       eprint = {2006.14663},
 primaryClass = {astro-ph.GA},
       adsurl = {https://ui.adsabs.harvard.edu/abs/2020ApJ...898...68B}
}

@ARTICLE{jones2018,
       author = {{Jones}, Michael G. and {Haynes}, Martha P. and {Giovanelli}, Riccardo and {Moorman}, Crystal},
        title = "{The ALFALFA H I mass function: a dichotomy in the low-mass slope and a locally suppressed `knee' mass}",
      journal = {\mnras},
         year = 2018,
        month = jun,
       volume = {477},
       number = {1},
        pages = {2-17},
          doi = {10.1093/mnras/sty521},
archivePrefix = {arXiv},
       eprint = {1802.00053},
 primaryClass = {astro-ph.GA},
       adsurl = {https://ui.adsabs.harvard.edu/abs/2018MNRAS.477....2J}
}

@ARTICLE{wyder2007,
       author = {{Wyder}, Ted K. and {Martin}, D. Christopher and {Schiminovich}, David and {Seibert}, Mark and {Budav{\'a}ri}, Tam{\'a}s and {Treyer}, Marie A. and {Barlow}, Tom A. and {Forster}, Karl and {Friedman}, Peter G. and {Morrissey}, Patrick and {Neff}, Susan G. and {Small}, Todd and {Bianchi}, Luciana and {Donas}, Jos{\'e} and {Heckman}, Timothy M. and {Lee}, Young-Wook and {Madore}, Barry F. and {Milliard}, Bruno and {Rich}, R. Michael and {Szalay}, Alex S. and {Welsh}, Barry Y. and {Yi}, Sukyoung K.},
        title = "{The UV-Optical Galaxy Color-Magnitude Diagram. I. Basic Properties}",
      journal = {\apjs},
         year = 2007,
        month = dec,
       volume = {173},
       number = {2},
        pages = {293-314},
          doi = {10.1086/521402},
archivePrefix = {arXiv},
       eprint = {0706.3938},
 primaryClass = {astro-ph},
       adsurl = {https://ui.adsabs.harvard.edu/abs/2007ApJS..173..293W}
}

@article{schlegel1998,
doi = {10.1086/305772},
url = {https://dx.doi.org/10.1086/305772},
year = {1998},
month = {jun},
publisher = {},
volume = {500},
number = {2},
pages = {525},
author = {Schlegel, David J. and Finkbeiner, Douglas P. and Davis, Marc},
title = {Maps of Dust Infrared Emission for Use in Estimation of Reddening and Cosmic Microwave Background Radiation Foregrounds},
journal = {The Astrophysical Journal},

}

@article{martin2005,
doi = {10.1086/426387},
url = {https://dx.doi.org/10.1086/426387},
year = 2005,
month = jan,
publisher = {},
volume = {619},
number = {1},
pages = {L1},
author = {Martin, D. Christopher and Fanson, James and Schiminovich, David and Morrissey, Patrick and Friedman, Peter G. and Barlow, Tom A. and Conrow, Tim and Grange, Robert and Jelinsky, Patrick N. and Milliard, Bruno and Siegmund, Oswald H. W. and Bianchi, Luciana and Byun, Yong-Ik and Donas, Jose and Forster, Karl and Heckman, Timothy M. and Lee, Young-Wook and Madore, Barry F. and Malina, Roger F. and Neff, Susan G. and Rich, R. Michael and Small, Todd and Surber, Frank and Szalay, Alex S. and Welsh, Barry and Wyder, Ted K.},
title = {The Galaxy Evolution Explorer: A Space Ultraviolet Survey Mission},
journal = {The Astrophysical Journal},

}

@article{mcquinn2015b,
doi = {10.1088/0004-637X/808/2/109},
url = {https://dx.doi.org/10.1088/0004-637X/808/2/109},
year = 2015,
month = jul,
publisher = {The American Astronomical Society},
volume = {808},
number = {2},
pages = {109},
author = {McQuinn, Kristen B. W. and Skillman, Evan D. and Dolphin, Andrew E. and Mitchell, Noah P.},
title = {CALIBRATING UV STAR FORMATION RATES FOR DWARF GALAXIES FROM STARBIRDS*},
journal = {The Astrophysical Journal},

}

@ARTICLE{Kennicutt1998a,
       author = {{Kennicutt}, Jr., Robert C.},
        title = "{Star Formation in Galaxies Along the Hubble Sequence}",
      journal = {\araa},
         year = 1998,
        month = jan,
       volume = {36},
        pages = {189-232},
          doi = {10.1146/annurev.astro.36.1.189},
archivePrefix = {arXiv},
       eprint = {astro-ph/9807187},
 primaryClass = {astro-ph},
       adsurl = {https://ui.adsabs.harvard.edu/abs/1998ARA&A..36..189K}
}

@article{Lee_2009,
doi = {10.1088/0004-637X/706/1/599},
url = {https://dx.doi.org/10.1088/0004-637X/706/1/599},
year = {2009},
month = {nov},
publisher = {The American Astronomical Society},
volume = {706},
number = {1},
pages = {599},
author = {Lee, Janice C. and Gil de Paz, Armando and Tremonti, Christy and Kennicutt, Robert C. and Salim, Samir and Bothwell, Matthew and Calzetti, Daniela and Dalcanton, Julianne and Dale, Daniel and Engelbracht, Chad and José G. Funes, S. J. and Johnson, Benjamin and Sakai, Shoko and Skillman, Evan and van Zee, Liese and Walter, Fabian and Weisz, Daniel},
title = {COMPARISON OF Hα AND UV STAR FORMATION RATES IN THE LOCAL VOLUME: SYSTEMATIC DISCREPANCIES FOR DWARF GALAXIES},
journal = {The Astrophysical Journal},
}

@article{Roychowdhury_2014,
    author = {Roychowdhury, Sambit and Chengalur, Jayaram N. and Kaisin, Serafim S. and Karachentsev, Igor D.},
    title = {The relation between atomic gas and star formation rate densities in faint dwarf irregular galaxies},
    journal = {Monthly Notices of the Royal Astronomical Society},
    volume = {445},
    number = {2},
    pages = {1392-1402},
    year = {2014},
    month = {10},
    issn = {0035-8711},
    doi = {10.1093/mnras/stu1814},
    url = {https://doi.org/10.1093/mnras/stu1814},
    eprint = {https://academic.oup.com/mnras/article-pdf/445/2/1392/18753863/stu1814.pdf},
}

@article{Walter_2008,
doi = {10.1088/0004-6256/136/6/2563},
url = {https://dx.doi.org/10.1088/0004-6256/136/6/2563},
year = {2008},
month = {nov},
publisher = {The American Astronomical Society},
volume = {136},
number = {6},
pages = {2563},
author = {Walter, Fabian and Brinks, Elias and de Blok, W. J. G. and Bigiel, Frank and Kennicutt, Robert C. and Thornley, Michele D. and Leroy, Adam},
title = {THINGS: THE H i NEARBY GALAXY SURVEY},
journal = {The Astronomical Journal},
}

@PHDTHESIS{masters_2005,
       author = {{Masters}, Karen Louise},
        title = "{Galaxy flows in and around the Local Supercluster}",
       school = {Cornell University, New York},
         year = 2005,
        month = jan,
       adsurl = {https://ui.adsabs.harvard.edu/abs/2005PhDT.........2M}
}

@MISC{marine_2023,
       author = {{Marine}, Joshua R.},
        title = "{Photometric Analysis of the Stellar Populations of 82 Low-Mass, Low Metallicity Galaxies using the Spitzer Space Telescope}",
       howpublished = {Undergraduate Honors Thesis, Macalester College, St. Paul, MN},
         year = 2023,
        month = jun,
        adsurl = {https://digitalcommons.macalester.edu/mjpa/vol11/iss1/10}
}

@MISC{shepley_2020,
       author = {{Shepley}, Madeline},
        title = "{Star Formation in Low-Mass Dwarf Galaxies}",
       howpublished = {Undergraduate Honors Thesis, Indiana University, Bloomington, IN},
         year = 2020,
        month = jun,
}

@article{Dale_2023,
doi = {10.3847/1538-3881/accffe},
url = {https://doi.org/10.3847/1538-3881/accffe},
year = {2023},
month = {may},
publisher = {The American Astronomical Society},
volume = {165},
number = {6},
pages = {260},
author = {Dale, Daniel A. and Boquien, Médéric and Turner, Jordan A. and Calzetti, Daniela and Kennicutt, Robert C. and Lee, Janice C.},
title = {Spectral Energy Distributions for 258 Local Volume Galaxies},
journal = {The Astronomical Journal},
}

@article{Hunter_2010,
doi = {10.1088/0004-6256/139/2/447},
url = {https://doi.org/10.1088/0004-6256/139/2/447},
year = {2010},
month = {jan},
publisher = {The American Astronomical Society},
volume = {139},
number = {2},
pages = {447},
author = {Hunter, Deidre A. and Elmegreen, Bruce G. and Ludka, Bonnie C.},
title = {GALEX ULTRAVIOLET IMAGING OF DWARF GALAXIES AND STAR FORMATION RATES*},
journal = {The Astronomical Journal},
}

@ARTICLE{Bell_2001,
       author = {{Bell}, Eric F. and {de Jong}, Roelof S.},
        title = "{Stellar Mass-to-Light Ratios and the Tully-Fisher Relation}",
      journal = {\apj},
         year = 2001,
        month = mar,
       volume = {550},
       number = {1},
        pages = {212-229},
          doi = {10.1086/319728},
archivePrefix = {arXiv},
       eprint = {astro-ph/0011493},
 primaryClass = {astro-ph},
       adsurl = {https://ui.adsabs.harvard.edu/abs/2001ApJ...550..212B}
}

@ARTICLE{salzer_1989,
       author = {{Salzer}, John J. and {MacAlpine}, Gordon M. and {Boroson}, Todd A.},
        title = "{Observations of a Complete Sample of Emission-Line Galaxies. I. CCD Imaging and Spectroscopy of Galaxies in UM Lists IV and V}",
      journal = {\apjs},
         year = 1989,
        month = jul,
       volume = {70},
        pages = {447},
          doi = {10.1086/191345},
       adsurl = {https://ui.adsabs.harvard.edu/abs/1989ApJS...70..447S}
}

@ARTICLE{bell&kennicutt2001,
       author = {{Bell}, Eric F. and {Kennicutt}, Jr., Robert C.},
        title = "{A Comparison of Ultraviolet Imaging Telescope Far-Ultraviolet and H{\ensuremath{\alpha}} Star Formation Rates}",
      journal = {\apj},
         year = 2001,
        month = feb,
       volume = {548},
       number = {2},
        pages = {681-693},
          doi = {10.1086/319025},
archivePrefix = {arXiv},
       eprint = {astro-ph/0010340},
 primaryClass = {astro-ph},
       adsurl = {https://ui.adsabs.harvard.edu/abs/2001ApJ...548..681B}
}

@ARTICLE{sullivan2000,
       author = {{Sullivan}, Mark and {Treyer}, Marie A. and {Ellis}, Richard S. and {Bridges}, Terry J. and {Milliard}, Bruno and {Donas}, Jos{\'e}},
        title = "{An ultraviolet-selected galaxy redshift survey - II. The physical nature of star formation in an enlarged sample}",
      journal = {\mnras},
         year = 2000,
        month = feb,
       volume = {312},
       number = {2},
        pages = {442-464},
          doi = {10.1046/j.1365-8711.2000.03140.x},
archivePrefix = {arXiv},
       eprint = {astro-ph/9910104},
 primaryClass = {astro-ph},
       adsurl = {https://ui.adsabs.harvard.edu/abs/2000MNRAS.312..442S}
}

@ARTICLE{salim2007,
       author = {{Salim}, Samir and {Rich}, R. Michael and {Charlot}, St{\'e}phane and {Brinchmann}, Jarle and {Johnson}, Benjamin D. and {Schiminovich}, David and {Seibert}, Mark and {Mallery}, Ryan and {Heckman}, Timothy M. and {Forster}, Karl and {Friedman}, Peter G. and {Martin}, D. Christopher and {Morrissey}, Patrick and {Neff}, Susan G. and {Small}, Todd and {Wyder}, Ted K. and {Bianchi}, Luciana and {Donas}, Jos{\'e} and {Lee}, Young-Wook and {Madore}, Barry F. and {Milliard}, Bruno and {Szalay}, Alex S. and {Welsh}, Barry Y. and {Yi}, Sukyoung K.},
        title = "{UV Star Formation Rates in the Local Universe}",
      journal = {\apjs},
         year = 2007,
        month = dec,
       volume = {173},
       number = {2},
        pages = {267-292},
          doi = {10.1086/519218},
archivePrefix = {arXiv},
       eprint = {0704.3611},
 primaryClass = {astro-ph},
       adsurl = {https://ui.adsabs.harvard.edu/abs/2007ApJS..173..267S}
}

@ARTICLE{weisz2012,
       author = {{Weisz}, Daniel R. and {Johnson}, Benjamin D. and {Johnson}, L. Clifton and {Skillman}, Evan D. and {Lee}, Janice C. and {Kennicutt}, Robert C. and {Calzetti}, Daniela and {van Zee}, Liese and {Bothwell}, Matthew S. and {Dalcanton}, Julianne J. and {Dale}, Daniel A. and {Williams}, Benjamin F.},
        title = "{Modeling the Effects of Star Formation Histories on H{\ensuremath{\alpha}} and Ultraviolet Fluxes in nearby Dwarf Galaxies}",
      journal = {\apj},
         year = 2012,
        month = jan,
       volume = {744},
       number = {1},
          eid = {44},
        pages = {44},
          doi = {10.1088/0004-637X/744/1/44},
archivePrefix = {arXiv},
       eprint = {1109.2905},
 primaryClass = {astro-ph.CO},
       adsurl = {https://ui.adsabs.harvard.edu/abs/2012ApJ...744...44W}
}

@ARTICLE{fumagalli2011,
       author = {{Fumagalli}, Michele and {da Silva}, Robert L. and {Krumholz}, Mark R.},
        title = "{Stochastic Star Formation and a (Nearly) Uniform Stellar Initial Mass Function}",
      journal = {\apjl},
         year = 2011,
        month = nov,
       volume = {741},
       number = {2},
          eid = {L26},
        pages = {L26},
          doi = {10.1088/2041-8205/741/2/L26},
archivePrefix = {arXiv},
       eprint = {1105.6101},
 primaryClass = {astro-ph.CO},
       adsurl = {https://ui.adsabs.harvard.edu/abs/2011ApJ...741L..26F}
}

@ARTICLE{dasilva2012,
       author = {{da Silva}, Robert L. and {Fumagalli}, Michele and {Krumholz}, Mark},
        title = "{SLUG{\textemdash}Stochastically Lighting Up Galaxies. I. Methods and Validating Tests}",
      journal = {\apj},
         year = 2012,
        month = feb,
       volume = {745},
       number = {2},
          eid = {145},
        pages = {145},
          doi = {10.1088/0004-637X/745/2/145},
archivePrefix = {arXiv},
       eprint = {1106.3072},
 primaryClass = {astro-ph.IM},
       adsurl = {https://ui.adsabs.harvard.edu/abs/2012ApJ...745..145D}
}

@ARTICLE{mcgaugh2014,
       author = {{Schombert}, James M. and {McGaugh}, Stacy},
        title = "{Stellar Populations and the Star Formation Histories of LSB Galaxies: IV Spitzer Surface Photometry of LSB Galaxies}",
      journal = {\pasa},
         year = 2014,
        month = feb,
       volume = {31},
          eid = {e011},
        pages = {e011},
          doi = {10.1017/pasa.2014.2},
archivePrefix = {arXiv},
       eprint = {1401.0238},
 primaryClass = {astro-ph.GA},
       adsurl = {https://ui.adsabs.harvard.edu/abs/2014PASA...31...11S}
}

@ARTICLE{lelli2016,
       author = {{Lelli}, Federico and {McGaugh}, Stacy S. and {Schombert}, James M.},
        title = "{SPARC: Mass Models for 175 Disk Galaxies with Spitzer Photometry and Accurate Rotation Curves}",
      journal = {\aj},
         year = 2016,
        month = dec,
       volume = {152},
       number = {6},
          eid = {157},
        pages = {157},
          doi = {10.3847/0004-6256/152/6/157},
archivePrefix = {arXiv},
       eprint = {1606.09251},
 primaryClass = {astro-ph.GA},
       adsurl = {https://ui.adsabs.harvard.edu/abs/2016AJ....152..157L}
}

@ARTICLE{TF2011,
       author = {{Torres-Flores}, S. and {Epinat}, B. and {Amram}, P. and {Plana}, H. and {Mendes de Oliveira}, C.},
        title = "{GHASP: an H{\ensuremath{\alpha}} kinematic survey of spiral and irregular galaxies - IX. The near-infrared, stellar and baryonic Tully-Fisher relations}",
      journal = {\mnras},
         year = 2011,
        month = sep,
       volume = {416},
       number = {3},
        pages = {1936-1948},
          doi = {10.1111/j.1365-2966.2011.19169.x},
archivePrefix = {arXiv},
       eprint = {1106.0505},
 primaryClass = {astro-ph.CO},
       adsurl = {https://ui.adsabs.harvard.edu/abs/2011MNRAS.416.1936T}
}

@ARTICLE{lelli2016b,
       author = {{Lelli}, Federico and {McGaugh}, Stacy S. and {Schombert}, James M.},
        title = "{The Small Scatter of the Baryonic Tully-Fisher Relation}",
      journal = {\apjl},
         year = 2016,
        month = jan,
       volume = {816},
       number = {1},
          eid = {L14},
        pages = {L14},
          doi = {10.3847/2041-8205/816/1/L14},
archivePrefix = {arXiv},
       eprint = {1512.04543},
 primaryClass = {astro-ph.GA},
       adsurl = {https://ui.adsabs.harvard.edu/abs/2016ApJ...816L..14L}
}

@ARTICLE{mcgaugh2012,
       author = {{McGaugh}, Stacy S.},
        title = "{The Baryonic Tully-Fisher Relation of Gas-rich Galaxies as a Test of {\ensuremath{\Lambda}}CDM and MOND}",
      journal = {\aj},
         year = 2012,
        month = feb,
       volume = {143},
       number = {2},
          eid = {40},
        pages = {40},
          doi = {10.1088/0004-6256/143/2/40},
archivePrefix = {arXiv},
       eprint = {1107.2934},
 primaryClass = {astro-ph.CO},
       adsurl = {https://ui.adsabs.harvard.edu/abs/2012AJ....143...40M}
}

@ARTICLE{cardamone2009,
       author = {{Cardamone}, Carolin and {Schawinski}, Kevin and {Sarzi}, Marc and {Bamford}, Steven P. and {Bennert}, Nicola and {Urry}, C.~M. and {Lintott}, Chris and {Keel}, William C. and {Parejko}, John and {Nichol}, Robert C. and {Thomas}, Daniel and {Andreescu}, Dan and {Murray}, Phil and {Raddick}, M. Jordan and {Slosar}, An{\v{z}}e and {Szalay}, Alex and {Vandenberg}, Jan},
        title = "{Galaxy Zoo Green Peas: discovery of a class of compact extremely star-forming galaxies}",
      journal = {\mnras},
         year = 2009,
        month = nov,
       volume = {399},
       number = {3},
        pages = {1191-1205},
          doi = {10.1111/j.1365-2966.2009.15383.x},
archivePrefix = {arXiv},
       eprint = {0907.4155},
 primaryClass = {astro-ph.CO},
       adsurl = {https://ui.adsabs.harvard.edu/abs/2009MNRAS.399.1191C}
}

@ARTICLE{yang2017,
       author = {{Yang}, Huan and {Malhotra}, Sangeeta and {Rhoads}, James E. and {Wang}, Junxian},
        title = "{Blueberry Galaxies: The Lowest Mass Young Starbursts}",
      journal = {\apj},
         year = 2017,
        month = sep,
       volume = {847},
       number = {1},
          eid = {38},
        pages = {38},
          doi = {10.3847/1538-4357/aa8809},
archivePrefix = {arXiv},
       eprint = {1706.02819},
 primaryClass = {astro-ph.GA},
       adsurl = {https://ui.adsabs.harvard.edu/abs/2017ApJ...847...38Y}
}

@ARTICLE{mcquinn2022,
      author = {{McQuinn}, Kristen. B.~W. and {Adams}, Elizabeth A.~K. and {Cannon}, John M. and {Fuson}, Jackson and {Skillman}, Evan D. and {Brooks}, Alyson and {Rhode}, Katherine L. and {Haynes}, Martha P. and {Inoue}, John L. and {Marine}, Joshua and {Salzer}, John. J. and {Talluri}, Anjana K.},
        title = "{The Turndown of the Baryonic Tully-Fisher Relation and Changing Baryon Fraction at Low Galaxy Masses}",
      journal = {\apj},
         year = 2022,
        month = nov,
       volume = {940},
       number = {1},
          eid = {8},
        pages = {8},
          doi = {10.3847/1538-4357/ac9285},
archivePrefix = {arXiv},
       eprint = {2203.10105},
 primaryClass = {astro-ph.GA},
       adsurl = {https://ui.adsabs.harvard.edu/abs/2022ApJ...940....8M}
}

@ARTICLE{sage1992,
       author = {{Sage}, L.~J. and {Salzer}, J.~J. and {Loose}, H.-H. and {Henkel}, C.},
        title = "{Star formation and molecular clouds in blue compact galaxies.}",
      journal = {\aap},
         year = 1992,
        month = nov,
       volume = {265},
        pages = {19},
       adsurl = {https://ui.adsabs.harvard.edu/abs/1992A&A...265...19S}
}

@ARTICLE{leroy2005,
       author = {{Leroy}, A. and {Bolatto}, A.~D. and {Simon}, J.~D. and {Blitz}, L.},
        title = "{The Molecular Interstellar Medium of Dwarf Galaxies on Kiloparsec Scales: A New Survey for CO in Northern, IRAS-detected Dwarf Galaxies}",
      journal = {\apj},
         year = 2005,
        month = jun,
       volume = {625},
       number = {2},
        pages = {763-784},
          doi = {10.1086/429578},
archivePrefix = {arXiv},
       eprint = {astro-ph/0502302},
 primaryClass = {astro-ph},
       adsurl = {https://ui.adsabs.harvard.edu/abs/2005ApJ...625..763L}
}

@ARTICLE{schruba2012,
       author = {{Schruba}, Andreas and {Leroy}, Adam K. and {Walter}, Fabian and {Bigiel}, Frank and {Brinks}, Elias and {de Blok}, W.~J.~G. and {Kramer}, Carsten and {Rosolowsky}, Erik and {Sandstrom}, Karin and {Schuster}, Karl and {Usero}, Antonio and {Weiss}, Axel and {Wiesemeyer}, Helmut},
        title = "{Low CO Luminosities in Dwarf Galaxies}",
      journal = {\aj},
         year = 2012,
        month = jun,
       volume = {143},
       number = {6},
          eid = {138},
        pages = {138},
          doi = {10.1088/0004-6256/143/6/138},
archivePrefix = {arXiv},
       eprint = {1203.4231},
 primaryClass = {astro-ph.CO},
       adsurl = {https://ui.adsabs.harvard.edu/abs/2012AJ....143..138S}
}

@article{Savino_2022,
doi = {10.3847/1538-4357/ac91cb},
url = {https://doi.org/10.3847/1538-4357/ac91cb},
year = {2022},
month = {oct},
publisher = {The American Astronomical Society},
volume = {938},
number = {2},
pages = {101},
author = {Savino, Alessandro and Weisz, Daniel R. and Skillman, Evan D. and Dolphin, Andrew and Kallivayalil, Nitya and Wetzel, Andrew and Anderson, Jay and Besla, Gurtina and Boylan-Kolchin, Michael and Bullock, James S. and Cole, Andrew A. and Collins, Michelle L. M. and Cooper, M. C. and Deason, Alis J. and Dotter, Aaron L. and Fardal, Mark and Ferguson, Annette M. N. and Fritz, Tobias K. and Geha, Marla C. and Gilbert, Karoline M. and Guhathakurta, Puragra and Ibata, Rodrigo and Irwin, Michael J. and Jeon, Myoungwon and Kirby, Evan and Lewis, Geraint F. and Mackey, Dougal and Majewski, Steven R. and Martin, Nicolas and McConnachie, Alan and Patel, Ekta and Rich, R. Michael and Simon, Joshua D. and Sohn, Sangmo Tony and Tollerud, Erik J. and van der Marel, Roeland P.},
title = {The Hubble Space Telescope Survey of M31 Satellite Galaxies. I. RR Lyrae–based Distances and Refined 3D Geometric Structure},
journal = {The Astrophysical Journal},
}

@article{Taylor_1998,
doi = {10.1086/300655},
url = {https://doi.org/10.1086/300655},
year = {1998},
month = {dec},
publisher = {},
volume = {116},
number = {6},
pages = {2746},
author = {Taylor, Christopher L. and Kobulnicky, Henry A. and Skillman, Evan D.},
title = {CO Emission in Low-Luminosity, H I-rich
Galaxies},
journal = {The Astronomical Journal}
}

@ARTICLE{dey_2019,
       author = {{Dey}, Arjun and {Schlegel}, David J. and {Lang}, Dustin and {Blum}, Robert and {Burleigh}, Kaylan and {Fan}, Xiaohui and {Findlay}, Joseph R. and {Finkbeiner}, Doug and {Herrera}, David and {Juneau}, St{\'e}phanie and {Landriau}, Martin and {Levi}, Michael and {McGreer}, Ian and {Meisner}, Aaron and {Myers}, Adam D. and {Moustakas}, John and {Nugent}, Peter and {Patej}, Anna and {Schlafly}, Edward F. and {Walker}, Alistair R. and {Valdes}, Francisco and {Weaver}, Benjamin A. and {Y{\`e}che}, Christophe and {Zou}, Hu and {Zhou}, Xu and {Abareshi}, Behzad and {Abbott}, T.~M.~C. and {Abolfathi}, Bela and {Aguilera}, C. and {Alam}, Shadab and {Allen}, Lori and {Alvarez}, A. and {Annis}, James and {Ansarinejad}, Behzad and {Aubert}, Marie and {Beechert}, Jacqueline and {Bell}, Eric F. and {BenZvi}, Segev Y. and {Beutler}, Florian and {Bielby}, Richard M. and {Bolton}, Adam S. and {Brice{\~n}o}, C{\'e}sar and {Buckley-Geer}, Elizabeth J. and {Butler}, Karen and {Calamida}, Annalisa and {Carlberg}, Raymond G. and {Carter}, Paul and {Casas}, Ricard and {Castander}, Francisco J. and {Choi}, Yumi and {Comparat}, Johan and {Cukanovaite}, Elena and {Delubac}, Timoth{\'e}e and {DeVries}, Kaitlin and {Dey}, Sharmila and {Dhungana}, Govinda and {Dickinson}, Mark and {Ding}, Zhejie and {Donaldson}, John B. and {Duan}, Yutong and {Duckworth}, Christopher J. and {Eftekharzadeh}, Sarah and {Eisenstein}, Daniel J. and {Etourneau}, Thomas and {Fagrelius}, Parker A. and {Farihi}, Jay and {Fitzpatrick}, Mike and {Font-Ribera}, Andreu and {Fulmer}, Leah and {G{\"a}nsicke}, Boris T. and {Gaztanaga}, Enrique and {George}, Koshy and {Gerdes}, David W. and {Gontcho}, Satya Gontcho A. and {Gorgoni}, Claudio and {Green}, Gregory and {Guy}, Julien and {Harmer}, Diane and {Hernandez}, M. and {Honscheid}, Klaus and {Huang}, Lijuan Wendy and {James}, David J. and {Jannuzi}, Buell T. and {Jiang}, Linhua and {Joyce}, Richard and {Karcher}, Armin and {Karkar}, Sonia and {Kehoe}, Robert and {Kneib}, Jean-Paul and {Kueter-Young}, Andrea and {Lan}, Ting-Wen and {Lauer}, Tod R. and {Le Guillou}, Laurent and {Le Van Suu}, Auguste and {Lee}, Jae Hyeon and {Lesser}, Michael and {Perreault Levasseur}, Laurence and {Li}, Ting S. and {Mann}, Justin L. and {Marshall}, Robert and {Mart{\'\i}nez-V{\'a}zquez}, C.~E. and {Martini}, Paul and {du Mas des Bourboux}, H{\'e}lion and {McManus}, Sean and {Meier}, Tobias Gabriel and {M{\'e}nard}, Brice and {Metcalfe}, Nigel and {Mu{\~n}oz-Guti{\'e}rrez}, Andrea and {Najita}, Joan and {Napier}, Kevin and {Narayan}, Gautham and {Newman}, Jeffrey A. and {Nie}, Jundan and {Nord}, Brian and {Norman}, Dara J. and {Olsen}, Knut A.~G. and {Paat}, Anthony and {Palanque-Delabrouille}, Nathalie and {Peng}, Xiyan and {Poppett}, Claire L. and {Poremba}, Megan R. and {Prakash}, Abhishek and {Rabinowitz}, David and {Raichoor}, Anand and {Rezaie}, Mehdi and {Robertson}, A.~N. and {Roe}, Natalie A. and {Ross}, Ashley J. and {Ross}, Nicholas P. and {Rudnick}, Gregory and {Safonova}, Sasha and {Saha}, Abhijit and {S{\'a}nchez}, F. Javier and {Savary}, Elodie and {Schweiker}, Heidi and {Scott}, Adam and {Seo}, Hee-Jong and {Shan}, Huanyuan and {Silva}, David R. and {Slepian}, Zachary and {Soto}, Christian and {Sprayberry}, David and {Staten}, Ryan and {Stillman}, Coley M. and {Stupak}, Robert J. and {Summers}, David L. and {Sien Tie}, Suk and {Tirado}, H. and {Vargas-Maga{\~n}a}, Mariana and {Vivas}, A. Katherina and {Wechsler}, Risa H. and {Williams}, Doug and {Yang}, Jinyi and {Yang}, Qian and {Yapici}, Tolga and {Zaritsky}, Dennis and {Zenteno}, A. and {Zhang}, Kai and {Zhang}, Tianmeng and {Zhou}, Rongpu and {Zhou}, Zhimin},
        title = "{Overview of the DESI Legacy Imaging Surveys}",
      journal = {\aj},
         year = 2019,
        month = may,
       volume = {157},
       number = {5},
          eid = {168},
        pages = {168},
          doi = {10.3847/1538-3881/ab089d},
archivePrefix = {arXiv},
       eprint = {1804.08657},
 primaryClass = {astro-ph.IM},
       adsurl = {https://ui.adsabs.harvard.edu/abs/2019AJ....157..168D}
}

@ARTICLE{jclee2009,
       author = {{Lee}, Janice C. and {Kennicutt}, Jr., Robert C. and {Funes}, S.~J. Jos{\'e} G. and {Sakai}, Shoko and {Akiyama}, Sanae},
        title = "{Dwarf Galaxy Starburst Statistics in the Local Volume}",
      journal = {\apj},
         year = 2009,
        month = feb,
       volume = {692},
       number = {2},
        pages = {1305-1320},
          doi = {10.1088/0004-637X/692/2/1305},
archivePrefix = {arXiv},
       eprint = {0810.5132},
 primaryClass = {astro-ph},
       adsurl = {https://ui.adsabs.harvard.edu/abs/2009ApJ...692.1305L}
}

@INPROCEEDINGS{skillman1987,
       author = {{Skillman}, Evan D.},
        title = "{Neutral Hydrogen and Star Formation in Irregular Galaxies}",
    booktitle = {NASA Conference Publication},
         year = 1987,
       editor = {{Lonsdale Persson}, Carol J.},
       series = {NASA Conference Publication},
       volume = {2466},
        month = may,
        pages = {263-266},
       adsurl = {https://ui.adsabs.harvard.edu/abs/1987NASCP2466..263S}
}

@ARTICLE{kennicutt1998sl,
       author = {{Kennicutt}, Jr., Robert C.},
        title = "{The Global Schmidt Law in Star-forming Galaxies}",
      journal = {\apj},
         year = 1998,
        month = may,
       volume = {498},
       number = {2},
        pages = {541-552},
          doi = {10.1086/305588},
archivePrefix = {arXiv},
       eprint = {astro-ph/9712213},
 primaryClass = {astro-ph},
       adsurl = {https://ui.adsabs.harvard.edu/abs/1998ApJ...498..541K}
}

@ARTICLE{roychowdhury2011,
       author = {{Roychowdhury}, Sambit and {Chengalur}, Jayaram N. and {Kaisin}, Serafim S. and {Begum}, Ayesha and {Karachentsev}, Igor D.},
        title = "{Small Bites: star formation recipes in extreme dwarfs}",
      journal = {\mnras},
         year = 2011,
        month = jun,
       volume = {414},
       number = {1},
        pages = {L55-L59},
          doi = {10.1111/j.1745-3933.2011.01055.x},
archivePrefix = {arXiv},
       eprint = {1103.6117},
 primaryClass = {astro-ph.CO},
       adsurl = {https://ui.adsabs.harvard.edu/abs/2011MNRAS.414L..55R}
}

@ARTICLE{begum2008a,
       author = {{Begum}, Ayesha and {Chengalur}, Jayaram N. and {Karachentsev}, I.~D. and {Sharina}, M.~E.},
        title = "{Baryonic Tully-Fisher relation for extremely low mass Galaxies}",
      journal = {\mnras},
         year = 2008,
        month = may,
       volume = {386},
       number = {1},
        pages = {138-144},
          doi = {10.1111/j.1365-2966.2008.13010.x},
archivePrefix = {arXiv},
       eprint = {0801.3606},
 primaryClass = {astro-ph},
       adsurl = {https://ui.adsabs.harvard.edu/abs/2008MNRAS.386..138B}
}

%%\allauthors


\end{document}